\documentclass[fleqn,usenatbib]{mnras}

\usepackage{newtxtext,newtxmath}
\usepackage[T1]{fontenc}

\DeclareRobustCommand{\VAN}[3]{#2}
\let\VANthebibliography\thebibliography
\def\thebibliography{\DeclareRobustCommand{\VAN}[3]{##3}\VANthebibliography}

\usepackage{graphicx}	
\usepackage{amsmath}	
\usepackage{subfig}     
\usepackage{pdfpages}   

\title[Short and long term activity analysis of GJ 436]{A Study of the Magnetic Activity and Variability of GJ 436}

\author[M. Kumar \& R. Fares]{
M. Kumar$^{1}$
and
R. Fares$^{1}$\thanks{Email:  \href{mailto:rim.fares@uaeu.ac.ae}{rim.fares@uaeu.ac.ae}}
\\
$^{1}$Department of Physics, College of Science, 
   United Arab Emirates University, P.O. Box No. 15551, Al Ain, UAE\\
}

\date{Accepted XXX. Received YYY; in original form ZZZ}

\pubyear{2022}

\begin{document}
\label{firstpage}
\pagerange{\pageref{firstpage}--\pageref{lastpage}}
\maketitle


\begin{abstract}
We present a magnetic activity study of GJ 436 using spectroscopic data from HARPS, spanning over 14 years, and additional data from NARVAL, falling within the HARPS observations. We study the CaII H\&K, HeID3, NaI doublet, H$\alpha$ and CaII IRT triplets lines and explore linear correlations between them. Using the full HARPS dataset, we found indices H$\alpha$ vs CaII H\&K \& H$\alpha$ vs HeI to correlate positively. From the NARVAL dataset, covering one observing epoch, we found CaII IRT$_{3}$ vs CaII IRT$_{2}$ \& CaII IRT$_{3}$ vs H$\alpha$ index to correlate negatively. We investigate long and short-term periodicity in these index variations using the Generalised Lomb-Scargle periodogram. For CaII H\&K, NaI and H$\alpha$ indices, we detect long-term periods of 2470.7d ($\approx$ 6.8 years), 1861.6d ($\approx$ 5.1 years) and 2160.9d ($\approx$ 5.9 years) respectively, consistent with GJ 436's photometric cycle of $\approx$ 7.4 years. Applying the "Pooled Variance" technique to H$\alpha$ \& NaI indices, we found $\approx$ 2500d to be the period of an activity cycle mechanism, in good agreement with the detected 2470.7d period. For CaII H\&K and H$\alpha$ indices, we detect short-term periods of $39.47^{+0.11}_{-0.15}$d and $40.46^{+0.44}_{-0.52}$d respectively, identified as the stellar rotation period. The stellar rotation is detected after prewhitening the long-term periodicity. It is detected as well in the analysis of individual observing epochs.

\end{abstract}




\section{Introduction} \label{Intro_sec}

M dwarfs, forming more than 75\% of main sequence stars in the solar neighbourhood \citep{2006AJ....132.2360H}, are ideal targets to search for low mass, habitable exoplanets. Due to their low luminosity, the habitable zone around them is close to their host star. In addition, their low mass and small size make them suitable for exoplanet detection \citep{2013ApJ...767...95D} using both the radial velocity and the transit methods. \par

M dwarfs, however, exhibit a different level of magnetic activity, which could affect the detection of exoplanets around them. Their activity manifests itself from the photosphere all the way to the corona, and thus can be studied from typical chromospheric activity indices to flares \citep[e.g.,][]{2022MNRAS.tmp.1043M, 2021tsc2.confE..18R, 2020A&A...637A..22R, 2020ApJ...892..144R,2019MNRAS.489..437D, 2019A&A...621A.126D, 2017ApJ...849...36Y, 2017A&A...600A..13A, 2010ApJ...710..924R}. Large-scale magnetic fields of M dwarfs have been studied thanks to spectropolarimetric campaigns \citep[e.g.,][]{2021A&ARv..29....1K,2012EAS....57..165M,2010MNRAS.407.2269M, 2008MNRAS.390..567M,2008MNRAS.390..545D}. In fact, the Rotation-Activity relationship shows that M dwarfs activity increases with the decrease of the rotation period, until they reach the saturation regime \citep{2017ApJ...834...85N, 2014ApJ...794..144R}. The large-scale magnetic field strength and topology varies with stellar mass and rotation. Fully convective M dwarfs have different properties from the partially convective ones. Large-scale magnetic fields, stellar wind and flares affect the environment at planetary orbits, as well as planetary emissions and atmospheres \citep{2022AN....34310079B,2021MNRAS.502.6201B,2021MNRAS.500L...1A, 2020MNRAS.494.2417V,2019ApJ...881..114Y,2019MNRAS.485.4529K, 2017MNRAS.470.4026V,2015MNRAS.449.4117V,2010MNRAS.406..409F} \par

Stellar activity is a major source of noise in exoplanets detection \citep{2016MNRAS.461.1465H,2012A&A...541A...9G,2011A&A...535A..55D}. If the star exhibits an activity cycle, its activity level changes with time. This makes the exoplanet radial velocity (RV) detection a real challenge, since the cycle affects RV measurements and could sometimes even conceal the RV signal \citep{2021A&A...652A..28L,2018A&A...612A..89S,2016A&A...595A..12S}. Studying and modelling stellar activity is thus a key element to detect small, earth-like planets \citep{2021A&A...648A.103D, 2020arXiv200513386H, 2016MNRAS.457.3637H}. \cite{2016A&A...595A..12S}, studying photometric time series of a sample of early M-dwarfs with rotation periods varying between 16.8d - 61.3d, found that stars in their sample exhibit a mean activity cycle of 6.0 ± 2.9 years. \par

GJ 436 (Ross 905) is a planet-hosting M2.5V dwarf. The star, a faint $V$ = 10.61, has a stellar mass of 0.44 $M_{\odot}$. It is a slow rotator  with an estimated vsin($i_{\star}$) of $0.330^{+0.091}_{-0.066}$ km/s. It hosts a hot-Neptune GJ 436b (with 2 potential exoplanet candidates to be confirmed \citep{2014AcA....64..323M}). The hot-Neptune orbits the host star at a distance of 0.0285 AU ($\approx$ 14.7 $R_{\star}$) with an orbital period of 2.644d \footnote{See Table \ref{tab:params_table} for the entire list of stellar and planetary parameters.}. Because of the close proximity to the star, GJ 436b is found to have lost a substantial amount of its atmosphere and is trailed by a giant exospheric cloud composed mainly of hydrogen atoms \citep{2015Natur.522..459E, 2015A&A...582A..65B, 2014ApJ...786..132K}. This makes it the first evaporating hot-Neptune discovered to date and an exciting M dwarf to study. \par

Recently, this star has been found to exhibit a long-term magnetic activity cycle of $\approx$ 7.4 years detected from 14 years of ground-based photometry \citep{2018AJ....155...66L}. This period is in good agreement with the findings of \cite{2016A&A...595A..12S}. The star is found to have its NaI activity index variations, representative of the mid-to-lower chromosphere, strongly correlate with its RV \citep{2012A&A...541A...9G}. Hence, a signal induced by stellar activity in NaI indices will be present in the RV measurements too. But the effect of stellar activity is not necessarily the same for all activity indicators and to constructively investigate the origin of variability in RV measurements, it is crucial to understand these different effects \citep{2021A&A...652A..28L}. \par
In this paper, we study the chromospheric lines CaII H\&K, HeID3, NaI, H$\alpha$, and CaII infrared triplet (IRT) for the M dwarf GJ 436. We calculate activity indices using these lines and study the stellar rotation \& long-term activity cycle periods detected from their variability. Multiple activity indices are used in order to explore the differences in their long-term variations with respect to one another. We explore linear correlations between these activity indices and compare them to previous activity correlation studies. We make use of all publicly available spectroscopic data of GJ 436 from the HARPS archive containing multiple observing epochs. Additionally, we use spectroscopic data from the échelle spectropolarimeter NARVAL, with its observations filling a gap in the HARPS observations. \par 
The paper is organised as follows. In Section \ref{Obs_sec}, we describe the data used for our analysis, and in Section \ref{Act_indicies_sec} we describe the activity indices calculation procedures. Correlations between the indices are presented in Section \ref{Comp_bet_indicies_sec}, and their variability in Section \ref{Act_ind_periodicity_sec}. In Section \ref{Discussion_sec},  we summarise and discuss the results and we conclude in Section \ref{Conclusion_sec}.

\section{Observations} \label{Obs_sec}

\subsection{HARPS} 

We used a total of 192 high-resolution spectra of GJ 436 from the European Southern Observatory (ESO) archive \footnote{\url{http://archive.eso.org/wdb/wdb/adp/phase3_spectral/form}}, spanning over 14 years from 25th January 2006 to 22nd March 2020. The spectra were obtained using the High Accuracy Radial velocity Planet Searcher (HARPS) spectrograph \citep{2003Msngr.114...20M} installed at the 3.6m ESO telescope in La Silla Observatory (Chile). HARPS has a high spectral resolution of R = 115,000, and a wavelength coverage of 378 - 691 nm. HARPS reduced spectra were not continuum normalised by the reduction pipeline \footnote{\url{https://www.eso.org/sci/facilities/lasilla/instruments/harps/doc/DRS.pdf}} and their wavelengths refer to the Barycentric rest frame. 
The exposure times of these spectra ranged from 200s to 2000s. The median signal-to-noise ratio (SNR) per spectra ranged from 6.95 to 39.5. From these 192 observations, we note that 44 observations were taken in one single night on 10th May 2007 and were primarily used for investigating the Rossiter-McLaughlin effect \citep{2014A&A...572A..73L}.

\subsection{NARVAL}

In addition to the 192 HARPS spectra, we used 16 high-resolution spectropolarimetric observations of GJ 436 obtained using NARVAL \citep{2003EAS.....9..105A}, an échelle spectropolarimeter installed at the 2m telescope Bernard Lyot (TBL) at Pic du Midi observatory in southwest France. The data were accessed from the Polarbase database \citep{2014PASP..126..469P}. The observations spanned 85 days, from 16th March 2016 to 8th June 2016. 
NARVAL provides a complete optical wavelength coverage of 370 - 1000 nm and has a resolving power of $\approx$ 65,000 in the polarimetric mode. Spectropolarimteric observation consisted of 4 sub-exposures of 700s each, taken at different angles of the polarisation wave-plates. The intensity spectrum is calculated by different combinations of these sub-exposures.
\footnote{For more info on how the intensity profiles are computed, see \cite{1997MNRAS.291..658D}}. \par 
NARVAL offers the use of the fully automated data reduction package $\tt LIBRE-ESpRIT$ \citep{1997MNRAS.291..658D} to all its users. The package automatically reduces and normalizes the spectra to unit continuum with their wavelengths referring to the Heliocentric rest frame. The peak SNR for our spectra ranged from 201 - 289 around $\approx$ 768nm.

\begin{table}
    \renewcommand\thetable{1}
	\centering
	\caption{Stellar and planetary parameters of the GJ 436 system. The respective references are mentioned in the last column.}
	
	\label{tab:params_table}
	\begin{tabular}{ccc}
		\hline
		Parameter & Value & Ref.\\
		\hline
		\hline
		  & \textbf{Star} & \\
		\hline  
		$M_{\star}$ ($M_{\odot}$) & 0.441±0.009 & \cite{2021ApJS..255....8R}\\
		$R_{\star}$ ($R_{\odot}$) & 0.417±0.008 & \cite{2021ApJS..255....8R}\\
		distance (pc) & 9.76±0.01 & \cite{2018yCat.1345....0G}\\
		$P_{rot}$ (d) & 44.09±0.08 & \cite{2018Natur.553..477B}\\
		$T_{eff}$(K) & 3586.10±36.37 & \cite{2021ApJS..255....8R}\\
		Sp.T & M2.5V & \cite{2004ApJ...617..580B}\\
		$log_{g}$ & 4.84 & \cite{2021ApJS..255....8R}\\
		Fe/H (dex) & 0.099±0.078 & \cite{2021ApJS..255....8R}\\
		$i_{\star}$ (degree) & $39^{+13}_{-9}$ & \cite{2018Natur.553..477B}\\
		vsin($i_{\star}$)(km/s) & $0.330^{+0.091}_{-0.066}$ & \cite{2018Natur.553..477B}\\
		$v_{rad}$(km/s) & 9.609±0.001 &{\cite{2018A&A...616A...1G}}\\
		$m_{V}$ & 10.61±0.01 & \cite{2012yCat.1322....0Z}\\
		$B - V$ & 1.447 ± 0.28 & {\cite{2000A&A...355L..27H}}\\
		\hline
		  & \textbf{Planet} & \\
		\hline  
		$M_{p}$ ($M_{\oplus}$) & $25.4^{+2.1}_{-2.0}$ & \cite{2018Natur.553..477B}\\
		$R_{p}$ ($R_{\oplus}$) & 4.191±0.109 & \cite{2016MNRAS.459..789T}\\
		$T_{p}$ (JD) & 2455959 & \cite{2014AcA....64..323M}\\
		$T_{eq}$(K) & 686±10 & \cite{2016MNRAS.459..789T}\\
		P (d) & 2.644±0.001 &{\cite{2018A&A...609A.117T}}\\
		e & $0.152^{+0.009}_{-0.008}$ &{\cite{2018A&A...609A.117T}}\\
		a(AU) & 0.0285±0.0002 & \cite{2021ApJS..255....8R}\\
		\hline
	\end{tabular}
\end{table}

\begin{table}
    \renewcommand\thetable{2}
	\centering
	\caption{Different chromospheric lines used to calculate activity indices. The table lists, for each line, the line core ($\lambda_{line}$), the bandwidth centered on the line core ($\Delta \lambda_{line}$), the blue \& red reference continuum wavelengths (Blue Cont. \& Red Cont.) and their corresponding bandwidths ($\Delta \lambda_{blue}$ \& $\Delta \lambda_{red}$, respectively) in Å. For the CaII H line, both the square bandwidth (0.4Å) and the triangular bandwidth (1.09Å) are mentioned.}
	\label{tab:Chromospheric_lines_table}
	\resizebox{\columnwidth}{!}{%
	\begin{tabular}{cccccccc}
		\hline
		Line & $\lambda_{line}$ & $\Delta \lambda_{line}$ & Blue Cont. & $\Delta \lambda_{blue}$ & Red Cont. & $\Delta \lambda_{red}$\\
		\hline
		\hline
		CaII K & 3933.664 & 1.09 & 3901.07 & 20 & 4001.07 & 20\\
		CaII H & 3968.47 & 0.4/1.09 & 3901.07 & 20 & 4001.07 & 20\\
		HeID3 & 5875.62 & 0.4 & 5869 & 5.0 & 5881 & 5.0\\
		NaI D2 & 5889.95 & 1.0 & 5805.0 & 10 & 6090.0 & 20\\
		NaI D1 & 5895.92 & 1.0 & 5805.0 & 10 & 6090.0 & 20\\
		H$\alpha$ & 6562.808 & 1.6 & 6550.87 & 10.75 & 6580.31 & 8.75\\
		CaI & 6572.795 & 0.8 & 6550.87 & 10.75 & 6580.31 & 8.75\\
		CaII IRT$_{1}$ & 8498 & 1.0 & 8490 & 2.0 & 8509 & 2.0\\
		CaII IRT$_{2}$ & 8542 & 1.0 & 8530 & 2.0 & 8566 & 2.0\\
		CaII IRT$_{3}$ & 8662 & 1.0 & 8641 & 2.0 & 8678 & 2.0\\
		\hline
	\end{tabular}%
	}
\end{table}

\section{Activity Indices} \label{Act_indicies_sec}
In this section, we describe the methods used for calculating the magnetic activity indices of CaII H\&K, HeID3, NaID, H$\alpha$ and CaII IRT lines. Additionally, we calculate the CaI line index, which is insensitive to changes in the stellar activity. The CaI index serves as a control for the significance of variations in the H$\alpha$ index.We note that the CaII IRT index is calculated only for the NARVAL spectra, since the HARPS spectra do not cover this wavelength. \par
Each NARVAL spectrum is doppler shift corrected using the stellar radial velocity from table \ref{tab:params_table}. Whereas each HARPS spectrum is corrected using the radial velocity obtained from each of their cross correlation function (CCF) profiles. \par
The NARVAL spectra consists of 40 individual spectral orders overlapping with one another at their ends. For cases where two spectral orders overlap in the spectral region containing a chromospheric line we are studying, we chose the order with the higher SNR for our analysis. The HARPS spectra consists of 72 spectral orders stitched together to form the entire spectrum with none overlapping with one another. \par
For the HARPS dataset, the CaII H\&K indices are calculated using the open source python package $\tt ACTIN$ \citep{2018JOSS....3..667G}. ESO data currently does not provide the error on each flux value for HARPS \footnote{ESO Phase 3 Data Release Description: \url{https://www.eso.org/rm/api/v1/public/releaseDescriptions/72}}. Therefore, the error on each flux value is approximated as the photon noise, $\sigma_{PN}$ along with the CCD readout noise, $\sigma_{RON}$ obtained from the CCF profiles of each observation as  

\begin{equation}
    \sigma_{flux} = \sqrt{\sigma_{PN}^{2} + \sigma_{RON}^{2}}
\end{equation}

where $\sigma_{PN} = \sqrt{flux}$. \par

\begin{figure} 
\begin{tabular}{c}
\subfloat{\includegraphics[width = \columnwidth, height=3.5cm]{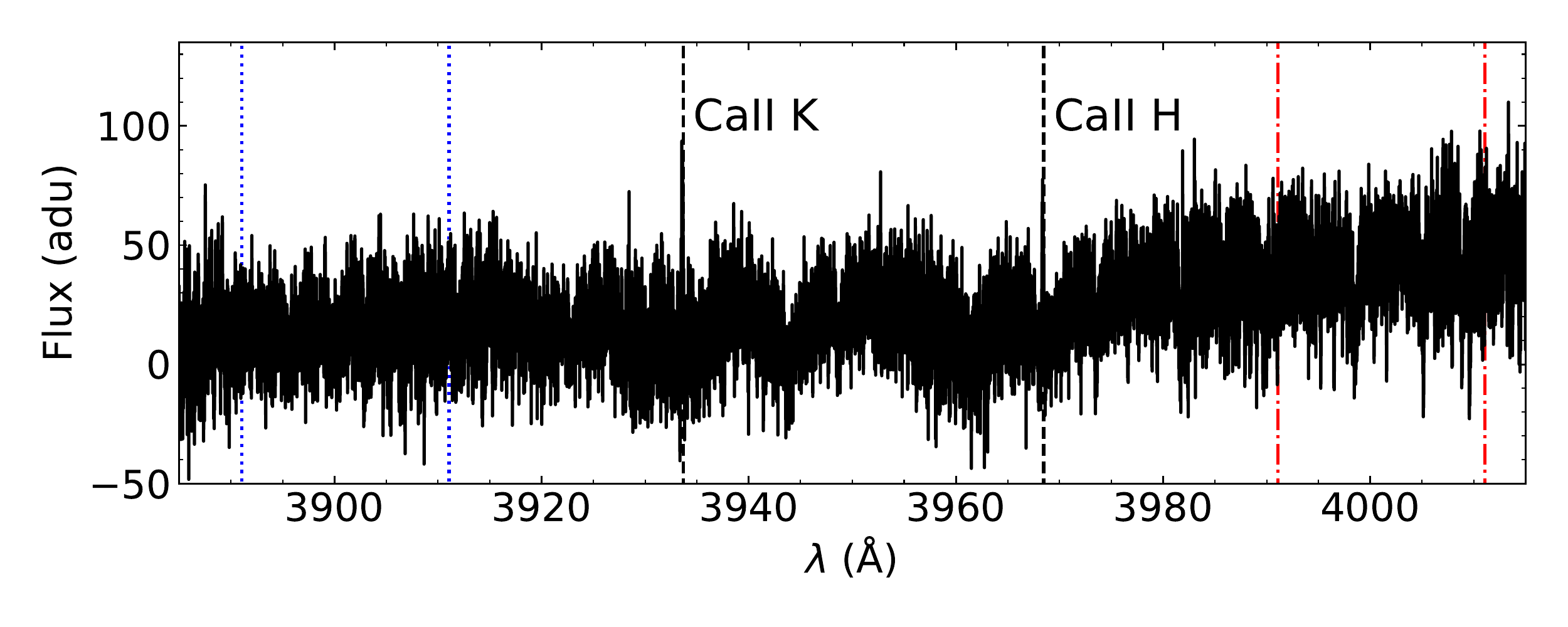}}\\

\subfloat{\includegraphics[width = \columnwidth, height=3.5cm]{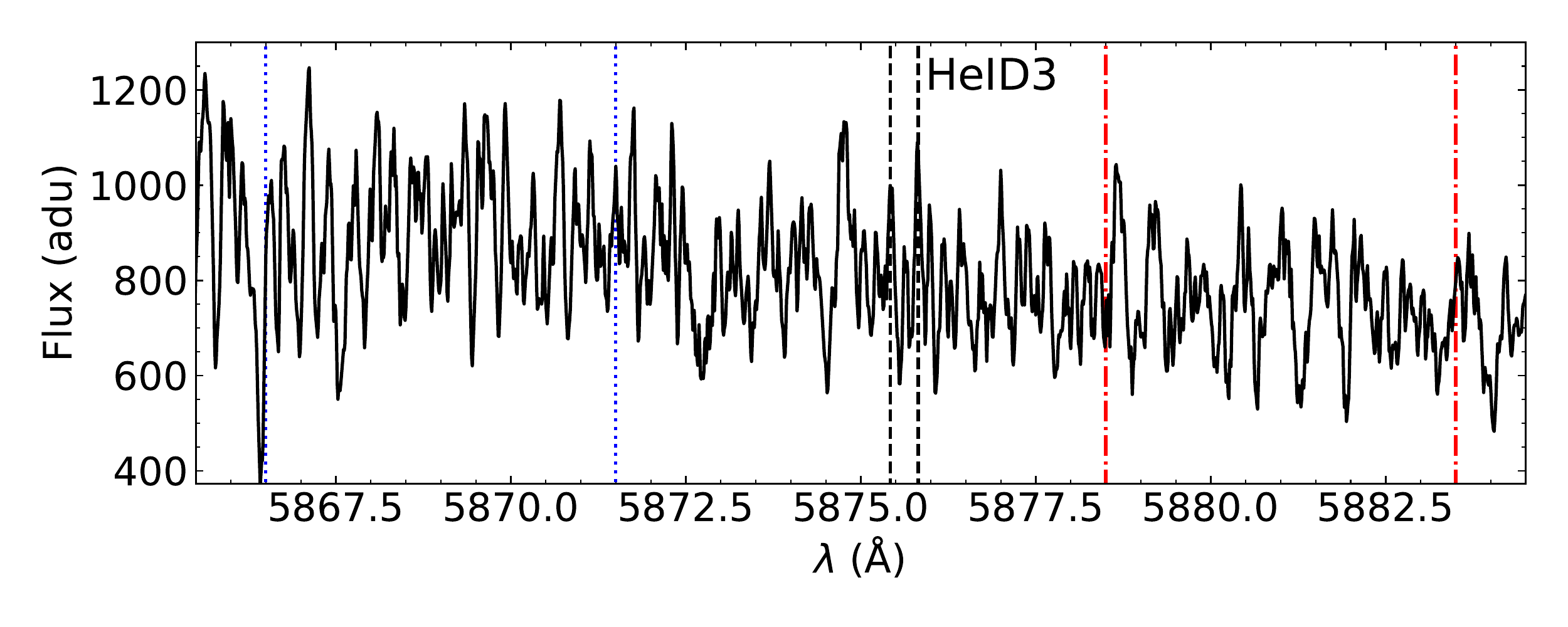}}\\

\subfloat{\includegraphics[width = \columnwidth, height=3.5cm]{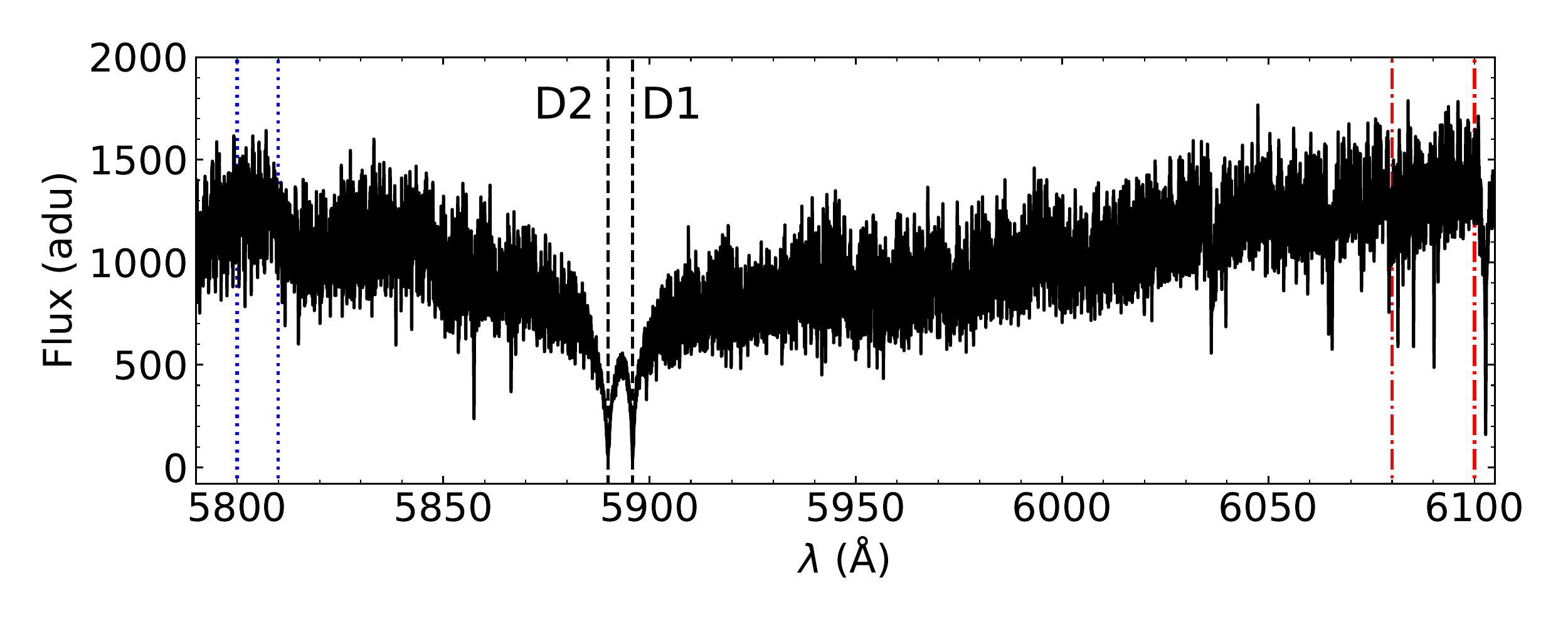}}\\

\subfloat{\includegraphics[width = \columnwidth, height=3.5cm]{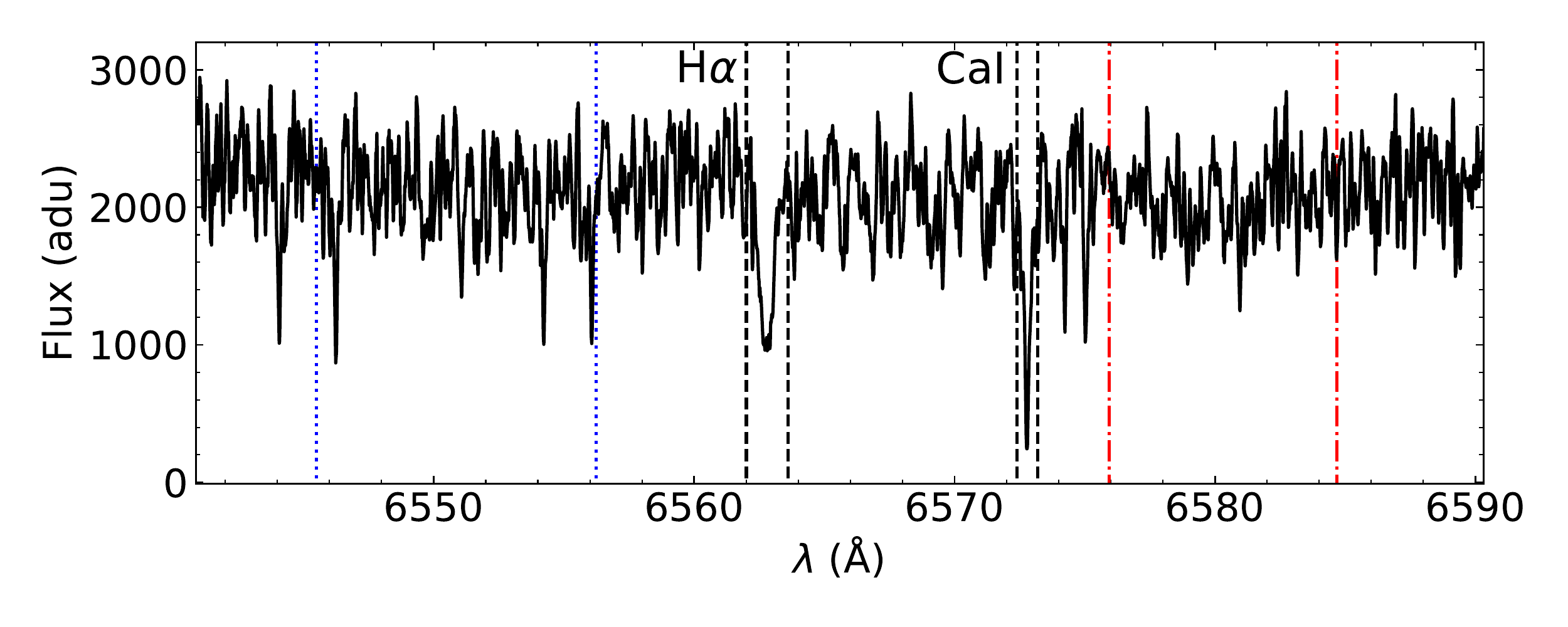}}\\

\end{tabular}
\caption{Region of the spectrum used to calculate activity indices for the CaII H\&K, HeI, NaI, H$\alpha$ \& CaI lines. The HARPS spectrum shown here was obtained on 25th January 2006. Within each figure, the blue and red vertical lines represent the reference continuum band used for each activity index. For HeI, H$\alpha$ \& CaI, the black vertical lines represent the line bandwidths used for index calculations. For CaII H\&K \& NaI, the black vertical lines represent the respective line cores. Bandwidths used for line cores and reference continuum for each index calculation are mentioned in table \ref{tab:Chromospheric_lines_table}.}
\label{fig:chromospheric_lines_plot}
\end{figure}

\begin{figure}
\begin{tabular}{c}
\subfloat{\includegraphics[width=\columnwidth, height=3.5cm]{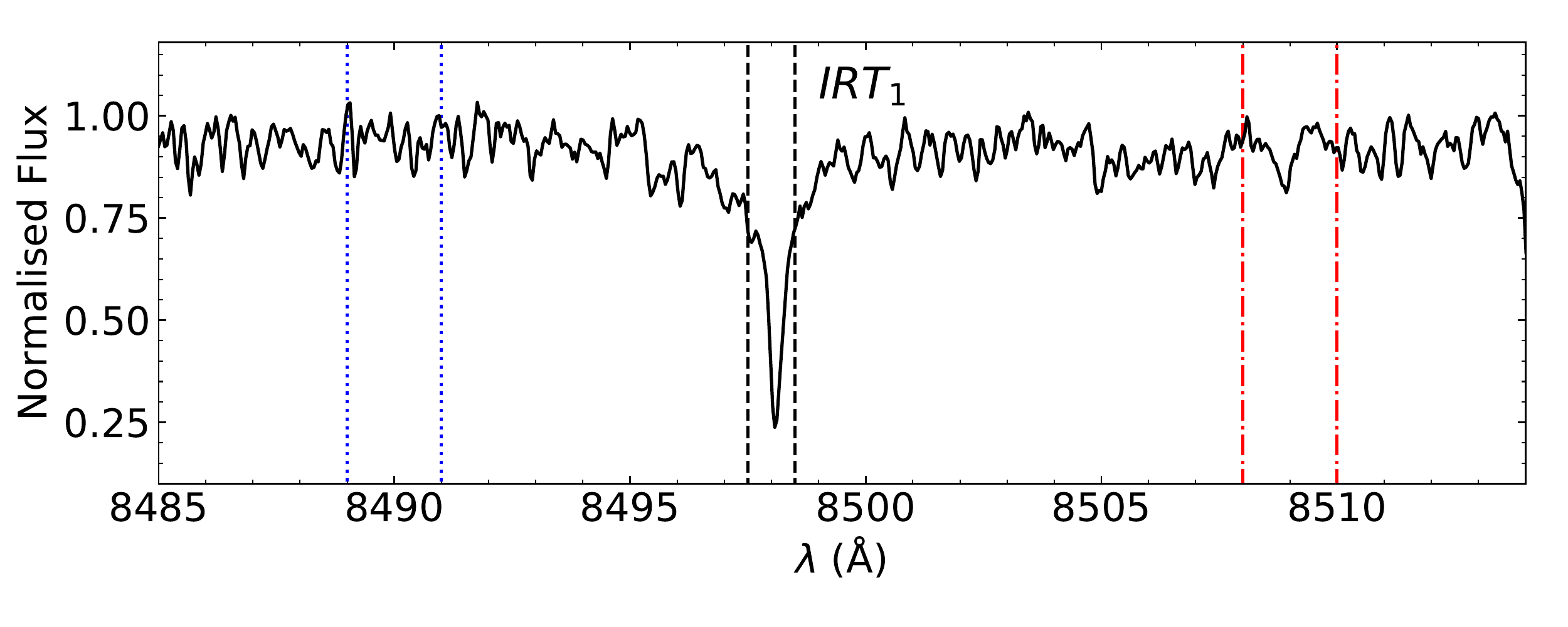}}\\
\subfloat{\includegraphics[width=\columnwidth, height=3.5cm]{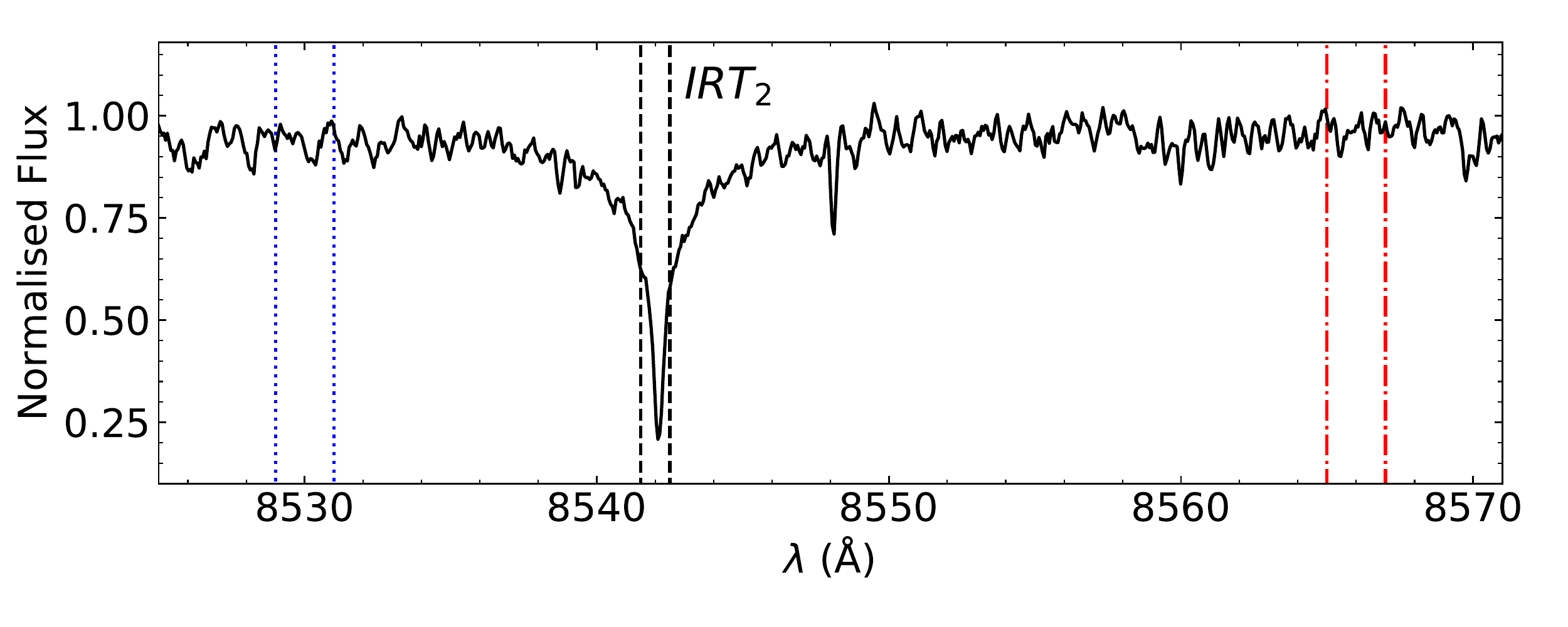}}\\
\subfloat{\includegraphics[width=\columnwidth, height=3.5cm]{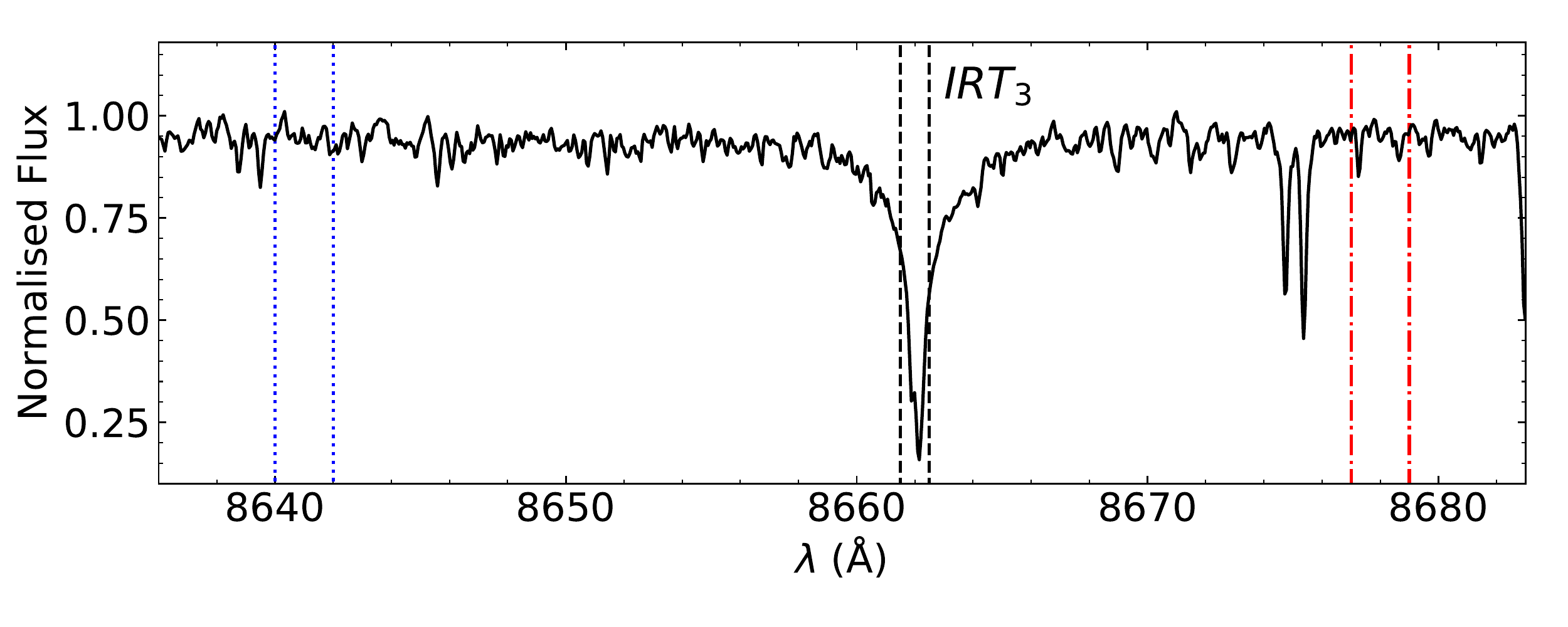}}\\
\end{tabular}
\caption{CaII IRT lines shown for one NARVAL spectrum obtained on 2nd June 2016. The blue and red vertical lines represent the reference continuum band and the black vertical lines represent the width of the line cores used for each IRT index calculation. Bandwidths used for these are detailed in table \ref{tab:Chromospheric_lines_table}.}

\label{fig:CaII_IRT_line_plot}
\end{figure}

The NARVAL dataset however has errors on each flux value provided by $\tt LIBRE-ESpRIT$. The errors on each of the indices are calculated using error propagation. The wavelength of each line, as well as the reference continuum used for the activity indices calculation for both datasets, are detailed in table \ref{tab:Chromospheric_lines_table}. Figure \ref{fig:chromospheric_lines_plot}  and \ref{fig:CaII_IRT_line_plot} shows the spectral region for each chromospheric line along with their respective reference continuum bands.

\subsection{CaII H\&K} \label{CaII_H_sec}
\label{sec:CaII}
The CaII H\&K lines are the most frequently explored magnetic activity proxy for solar-type stars \citep[see, e.g.,][]{1998ASPC..154..153B}. But for M dwarfs, their study could be challenging, since it requires longer exposure times due to their faint luminosity and low continuum levels \citep{2017A&A...598A..28S}. However, it provides significant information about the stellar chromosphere structure \citep{2020A&A...642A..53D}. \par 

There were two spectral orders in our NARVAL dataset containing one of each of the H and K emission lines. Our spectra around these lines were relatively noisy, compared to the H$\alpha$ line, with a mean SNR of $\approx$ 11. The reduced spectra obtained from LIBRE-ESpRIT were not normalised well. We applied a re-normalisation procedure of fitting a 4th order polynomial to the continuum, from which, the spectral order containing the CaII H line was well normalised but the one containing the CaII K line was poorly normalised. Even after removing a few strong photospheric lines, the normalisation was not reliable (see Appendix \ref{CaII HK_appendix_sec}). We thus decided to study only the CaII H line for our NARVAL spectra defining our so called $I_{CaII H}$ index following \cite{2012A&A...540A.138M} as 

\begin{equation} \label{CaII H_index_eq}
I_{CaII H} = \frac{F_{CaII H}}{F_{cont}}
\end{equation}
where $F_{CaII H}$ is the mean flux of the CaII H line within a 0.4Å wide rectangular window centered on 3968.47Å and $F_{cont}$ is the mean flux of the continuum within a 20Å wide rectangular window centered on 4001.07Å. The 0.4Å bandwidth was chosen as such to only encompass the CaII H emission line within it. \par

Since the continuum around the CaII H\&K lines was relatively less noisy in the HARPS spectra, we calculate its CaII H\&K index using $\tt ACTIN$ as 
\begin{equation} \label{CaII_index_eq}
I_{CaII} = \frac{F_{CaII K} + F_{CaII H}}{F_{1} + F_{2}}
\end{equation}

where $F_{CaII K}$ \& $F_{CaII H}$ are the mean fluxes within a 1.09Å triangular window centered on 3933.664Å and 3968.47Å respectively and $F_{1}$ \& $F_{2}$ are the mean fluxes within a 20Å square window centred on 3901.07Å and 4001.07Å respectively. 

\subsection{HeI D3} \label{HeID3_sec}
The HeI D3 line is observed to show presence of non-radiative heating mechanisms in cool stars \citep{1981ApJ...244..345L} and can be used to investigate both the existence and variation of magnetic activity \citep{1997A&A...326..741S}. The activity index is defined following \cite{2011A&A...534A..30G} as

\begin{equation} \label{HeI_index_eq}
I_{HeI} = \frac{F_{HeI}}{F_{1} + F_{2}}
\end{equation}

where $F_{HeI}$ is the mean flux within a 0.4Å wide bandwidth centered on the HeI D3 line 5875.62Å. $F_{1}$ \& $F_{2}$ are the mean fluxes of the blue and red continuum centered on 5869Å \& 5881Å and calculated within a bandwidth of 5.0Å respectively.

\subsection{NaI} \label{NaI_sec}
The NaI D resonance lines index is a good complement to the H$\alpha$ index since it represents the conditions of the middle-to-lower chromosphere as opposed to the upper chromosphere by H$\alpha$ \citep{2000ApJ...539..858M}. We define our index following \cite{2007MNRAS.378.1007D} as 
\begin{equation} \label{NaID_index_eq}
I_{NaI} = \frac{f_{1} + f_{2}}{f_{cont}}
\end{equation}
where $f_{1}$ \& $f_{2}$ are the mean fluxes within a 1Å wide bandwidth centered on the D1 5895.92Å and D2 5889.95Å line cores. $f_{cont}$ is defined as the pseudo-continuum calculated as
\begin{equation} \label{NaID_f_cont_eq}
f_{cont} = \frac{F_{1} + F_{2}}{2}
\end{equation}
where $F_{1}$ \& $F_{2}$ are the mean fluxes of the 10 highest flux values within two reference bands of width 10Å and 20Å centered on 5805Å and 6090Å respectively. \par 

\subsection{\texorpdfstring{H$\alpha$}{HaI}} \label{H_alpha_index_sec}
The H$\alpha$ line is sensitive to chromospheric activity and is extensively used as a standard activity tracer for M dwarfs \citep{1982ApJ...260..670L, 2004AJ....128..426W}. In both the NARVAL and HARPS spectra, the H$\alpha$ line is in absorption since GJ 436 is a low activity M dwarf. 

The standard approach of calculating the H$\alpha$ index is implemented following \cite{2009A&A...495..959B} as
\begin{equation} \label{H_alpha_index_eq}
I_{H\alpha} = \frac{F_{H\alpha}}{F_{1} + F_{2}}
\end{equation}
where $F_{H\alpha}$ is the mean flux within a broad 1.6Å wide bandwidth centered on the H$\alpha$ line 6562.808Å. $F_{1}$ \& $F_{2}$ are the mean fluxes of the blue and red continuum on either side of the line core within 10.75Å and 8.75Å wide bandwidths centered on 6550.87Å and 6580.31Å respectively.\par

\subsection{CaI} \label{CaI_sec}
The CaI line is observed a few angstroms to the right of the H$\alpha$ line at 6572.795Å and is known to not vary with stellar magnetic activity \citep{2003A&A...403.1077K}. Because of this, we calculate the CaI index as a check for the H$\alpha$ activity significance since any variation in H$\alpha$ activity should not be seen in the CaI indices. The index is defined following \cite{2013ApJ...764....3R} as
\begin{equation} \label{CaI_index_eq}
I_{CaI} = \frac{F_{CaI}}{F_{1} + F_{2}}
\end{equation}
where $F_{CaI}$ is the mean flux of the CaI line in a 0.8Å wide window centered on the CaI line and $F_{1}$ and $F_{2}$ are the mean fluxes of the same continuum as that used for the $I_{H\alpha}$ index calculation in Sect. \ref{H_alpha_index_sec}. The 0.8Å bandwidth was chosen to contain only the CaI line within it. The same spectral order used for the $I_{H\alpha}$ index calculation is used for calculating the CaI index. Each activity index time series figures are included in the supplementary material and are available online. \par

\subsection{CaII IRT} \label{CaII_IRT_sec}
The CaII IRT lines provide useful information about the middle chromosphere and can be used as an activity proxy for late-type main sequence stars \citep[see, e.g.,][]{2017A&A...607A..87M, 2007ASPC..368...27R}. The HARPS spectra do not cover the CaII IRT lines. On the other hand, the NARVAL spectra do, since the spectropolarimeter has a wavelength coverage of 370 - 1000nm. We calculate the CaII IRT index for the 16 NARVAL spectra obtained in 2016. \par
The index for each of the three IRT lines is defined following \cite{2017A&A...607A..87M} as
\begin{equation} \label{CaII_IRT_index_eq}
I_{IRT_x} = \frac{F_{IRT_x}}{F_{1} + F_{2}}
\end{equation}
where $F_{IRT_x}$ is the mean flux of the line, $F_{1}$ \& $F_{2}$ are the mean fluxes in the reference continuum bands. Here, $x$ = 1, 2, and 3, corresponds to each of the triplet lines at 8498Å, 8542Å \& 8662Å respectively (see table \ref{tab:Chromospheric_lines_table}). The $F_{IRT_x}$ mean fluxes are calculated within a 1Å wide bandwidth and the continuum mean fluxes within a 2Å wide bandwidth. \par

Table \ref{tab:Analysis_values_table} and \ref{tab:HARPS_analysis_table} contain the values for each calculated index along with their errors for the HARPS and NARVAL spectra respectively.

\begin{figure*}
    \includegraphics[width = 2\columnwidth]{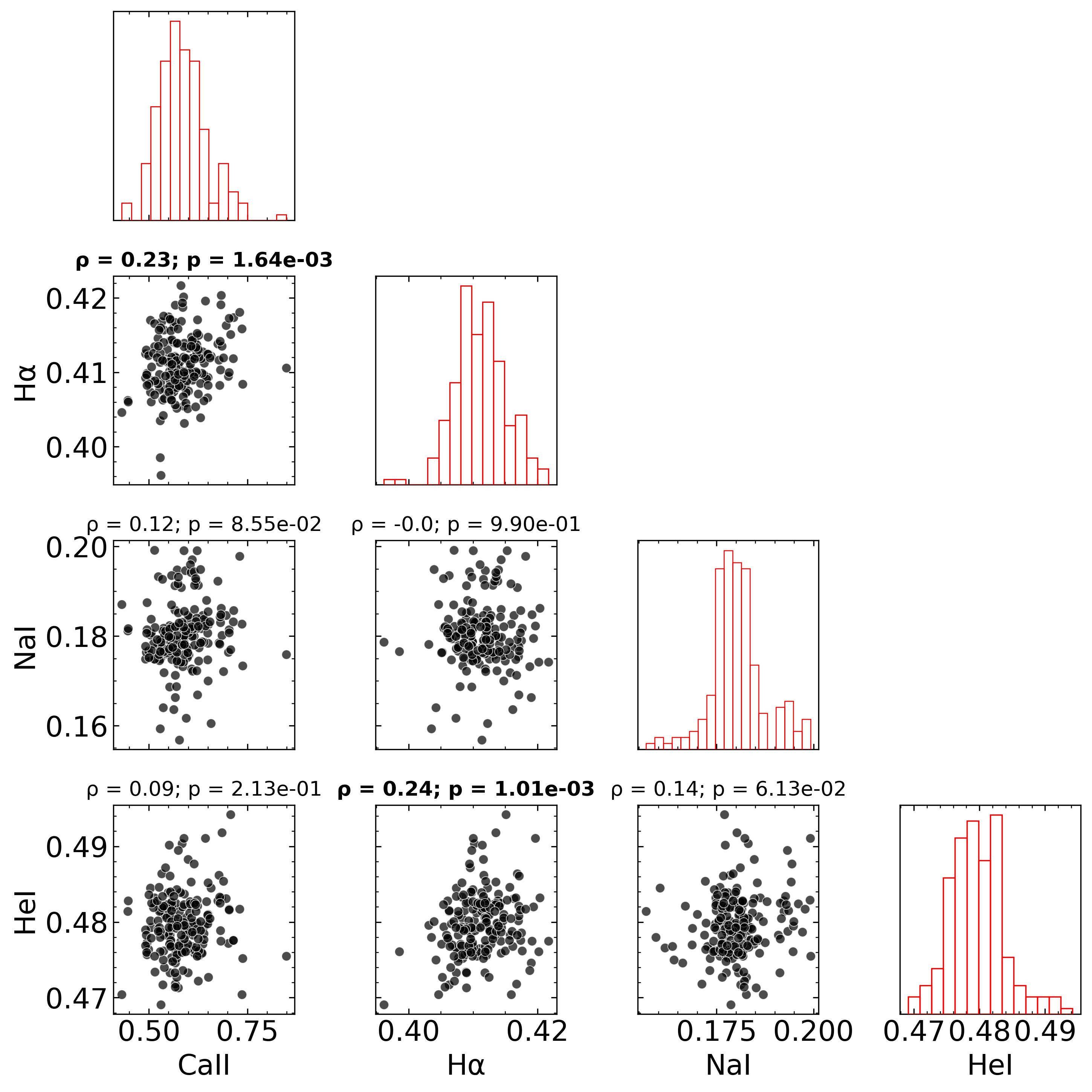}
    \caption{Corner plot showing correlations between the four activity indices CaII H\&K, HeI NaI, and H$\alpha$, for 192 HARPS spectra. The diagonal panels show the histogram distribution of each index. The Pearson R correlation coefficient ($\rho$) and its p-value are shown above each correlation plot, with significant correlation values shown in bold.}
    \label{fig:HARPS_full_corr_corner_plot}
\end{figure*}

\begin{table*}
    \renewcommand\thetable{5}
	\centering
	\caption{Pearson R correlation between activity indices per epoch. For each correlation, the shown results are [$\rho$, p-value, (slope)] where $\rho$ is the correlation coefficient, p-value is the corresponding null-hypothesis probability and slope is the slope of its best-fit line. Correlations with their p-value < 0.05 are shown in bold. For the 2016 (NARVAL) epoch, values are shown for correlations with CaII H instead of CaII H\&K.}
	\label{tab:pearson_correlation_per_epoch_table}
	\resizebox{2\columnwidth}{!}{%
	\begin{tabular}{ccccccc}
		\hline
		Epoch & H$\alpha$ vs CaII H\&K & H$\alpha$ vs NaI & NaI vs CaII H\&K & CaII H\&K vs HeI & H$\alpha$ vs HeI & NaI vs HeI\\
		\hline
		\hline
		
		2006 & 0.4506, 0.3103, (6.2344) & -0.6655, 0.1028, (-0.9041) & -0.2644, 0.5667, (-2.6923) & -0.5022, 0.2508, (-0.0546) & -0.4812, 0.2743, (-0.7233) & 0.4285, 0.3375, (0.4741)\\
		2007 & \textbf{0.3459, 0.0045, (7.4930)} & 0.1785, 0.1516, (0.3953) & 0.1066, 0.3945, (1.0424) & -0.0395, 0.7529, (-0.0026) & 0.1627, 0.1919, (0.2314) & 0.0299, 0.8117, (0.0192)\\
		2008 & 0.0921, 0.5380, (0.9013) & \textbf{-0.2896, 0.0483, (-0.3309)} & \textbf{0.2925, 0.0460, (2.5049)} & -0.0342, 0.8194, (-0.0031) & -0.1256, 0.4003, (-0.1129) & 0.2690, 0.0680, (0.2117)\\
		2009 & 0.1514, 0.3512, (1.7055) & 0.1718, 0.2891, (0.1974) & 0.0087, 0.9577, (0.0850) & 0.1208, 0.4580, (0.0125) & \textbf{0.5898, 0.00006, (0.6853)} & -0.0361, 0.8248, (-0.0365)\\
		2010 & \textbf{0.8977, 0.0010, (20.5098)} & -0.3035, 0.4273, (-0.3232) & -0.0264, 0.9462, (-0.5665) & 0.5106, 0.1601, (0.0289) & 0.6501, 0.0580, (0.8392) & \textbf{-0.6982, 0.0365, (-0.8462)}\\
		2016 & -0.1290, 0.7060, (-4.1041) & 0.1507, 0.5862, (0.3671) & \textbf{-0.612, 0.0117, (-7.9959)} & -0.3390, 0.1990, (-0.0184) & 0.1794, 0.5063, (0.3094) & 0.1850, 0.4927, (0.1311)\\
		2020 & 0.2235, 0.3054, (2.7487) & 0.0948, 0.6669, (0.0819) & -0.0151, 0.9454, (-0.2154) & 0.2237, 0.3049, (0.0211) & -0.0946, 0.6678, (-0.1098) & 0.0617, 0.7799, (0.0829)\\
		\hline
	\end{tabular}
	}
\end{table*}

\begin{figure*}
\begin{tabular}{cc}
\subfloat{\includegraphics[width = \columnwidth, height=3.5cm]{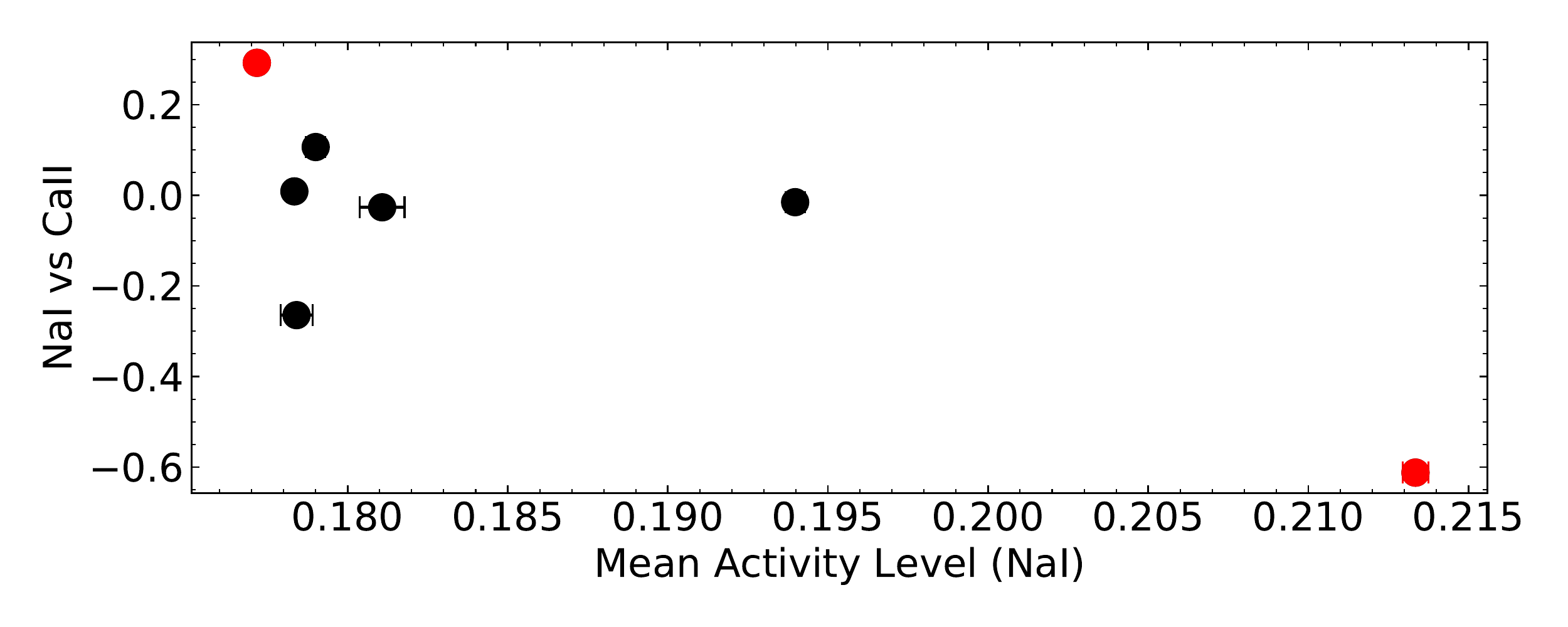}} &
\subfloat{\includegraphics[width = \columnwidth, height=3.5cm]{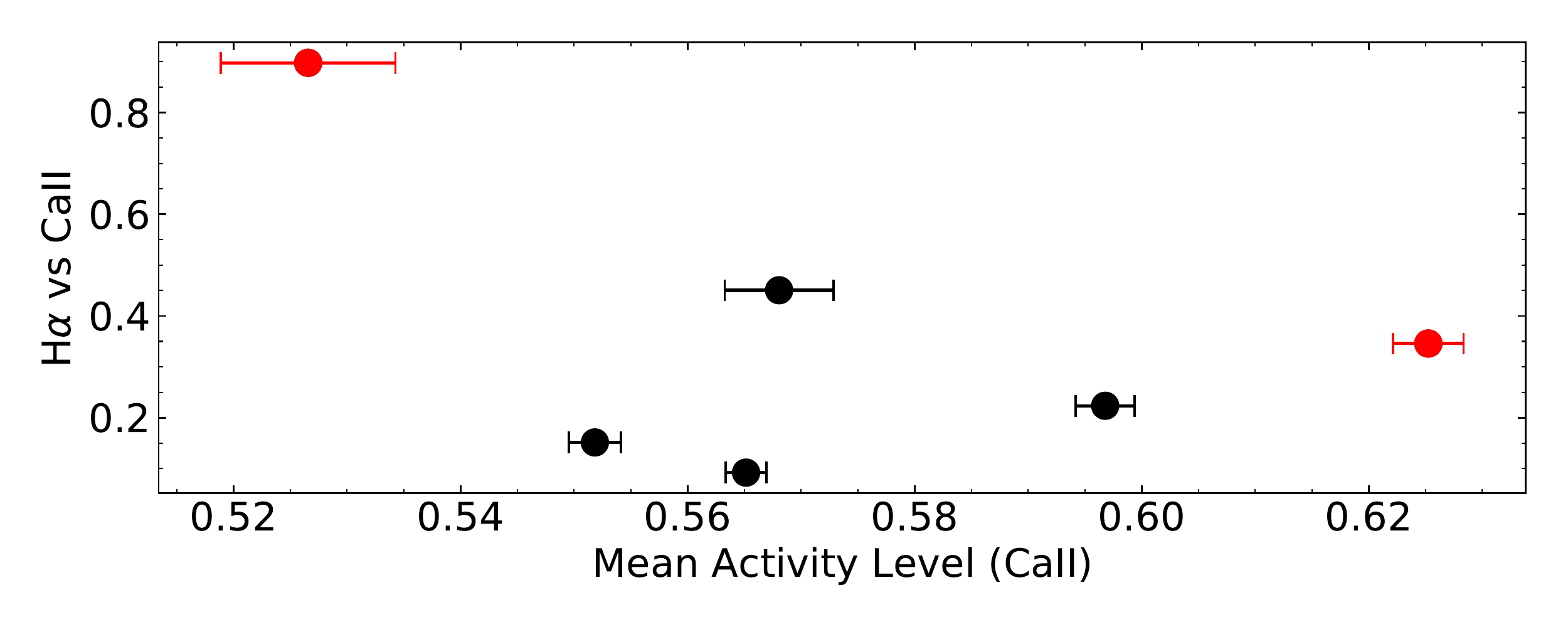}}\\

\end{tabular}
\caption{Correlation coefficients ($\rho$) plotted against the mean activity level per observing epoch for 2 pairs of index correlations. For H$\alpha$ vs CaII, we only plot the HARPS epoch correlations since we calculate a CaII H index for the NARVAL epoch. In each plot, the red dot markers are correlation coefficients with p-value < 0.05.}
\label{fig:corr_per_epoch_plots}
\end{figure*}

\section{Correlation between indices} \label{Comp_bet_indicies_sec}

In this section, we explore linear correlations between the CaII H\&K, HeI, NaI, CaII IRT, and H$\alpha$ indices, each representing the lower, mid-to-lower, middle and upper chromosphere respectively. Additionally, we compare the CaI line index to the H$\alpha$ index to see whether it correlates with an index insensitive to magnetic activity. The linear correlations are studied using the Pearson R correlation coefficient with correlations considered significant if their null-hypothesis probabilities, i.e.  p-values, are < 0.05. We remind the reader here that for the NARVAL dataset, the CaII H index was calculated instead of CaII H\&K. \par

\subsection{Correlations for HARPS dataset}

We begin by analysing the full HARPS dataset covering $\approx$ 14 years. Figure \ref{fig:HARPS_full_corr_corner_plot} shows a corner plot with correlations for each of the 4 activity indices. Each subplot has its respective Pearson R correlation coefficient and the p-value shown above it. \par 

We found the CaII H\&K indices to show a weak positive correlation with the H$\alpha$ indices with a coefficient of 0.23 (p=0.0016). Furthermore, the HeI index showed a significant positive correlation with the H$\alpha$ index with a coefficient of 0.24 (p=0.0010). \par 
The correlation between CaI and H$\alpha$ indices was found to be insignificant with a coefficient of -0.07 (p=0.318).This shows that the variations in H$\alpha$ indices are real, since the CaI line is insensitive to stellar activity. \par

\subsection{Correlations per Epoch}

In order to study how these correlations vary over long-term, and the effect of a possible cycle on them, we analyze separately the activity index correlations from each observing epoch of the HARPS dataset. Here, we treat the NARVAL dataset as an epoch along with the six HARPS epochs, with each having a mean timespan and number of data points of $\approx$ 90 days and 30 respectively. The correlation coefficients per epoch along with their p-values, are listed in table \ref{tab:pearson_correlation_per_epoch_table} for both NARVAL and HARPS dataset with significant correlations shown in bold. Corner plots showing correlations per observing epoch are included in the supplementary material and are available online. \par

Out of the 7 observing epochs, two epochs showed a significant correlation between the H$\alpha$ and CaII H\&K indices as shown in table \ref{tab:pearson_correlation_per_epoch_table}. The correlation coefficient in the 2010 epoch is roughly 2.6 times higher than that in the 2007 epoch. The NaI vs CaII H\&K index showed a significant correlation for 2 epochs as well. However, their correlation switches from positive, in 2008, to negative, in 2016. The absolute value of their correlation coefficient nearly doubles between these epochs. All of these results demonstrate that activity index correlations for GJ 436 vary with time. \par

In addition to these correlations, for the epoch of observations done with NARVAL, we found the CaII IRT indices to show significant correlations for IRT$_{2}$ vs IRT$_{3}$ \& H$\alpha$ vs IRT$_{3}$ with their correlation coefficients being -0.56 (p=0.024) \& -0.54 (p=0.031) respectively.

We further investigate whether the correlations change with the level of stellar activity. For that, we study the correlation coefficient variation as a function of the mean activity level (MAL) of each epoch, for the indices that show a significant correlation in at least two epochs of observation. The error on MAL is calculated using error propagation. As shown in figure \ref{fig:corr_per_epoch_plots}, the NaI vs CaII \& H$\alpha$ vs CaII index correlation coefficients tend to show a general trend of decreasing with increasing MAL of NaI \& CaII per epoch respectively. If only the significant correlations are considered, shown with the red dot markers, the trend remains the same.

\begin{figure*}
\begin{tabular}{cc}
\subfloat{\includegraphics[width = \columnwidth, height=3.5cm]{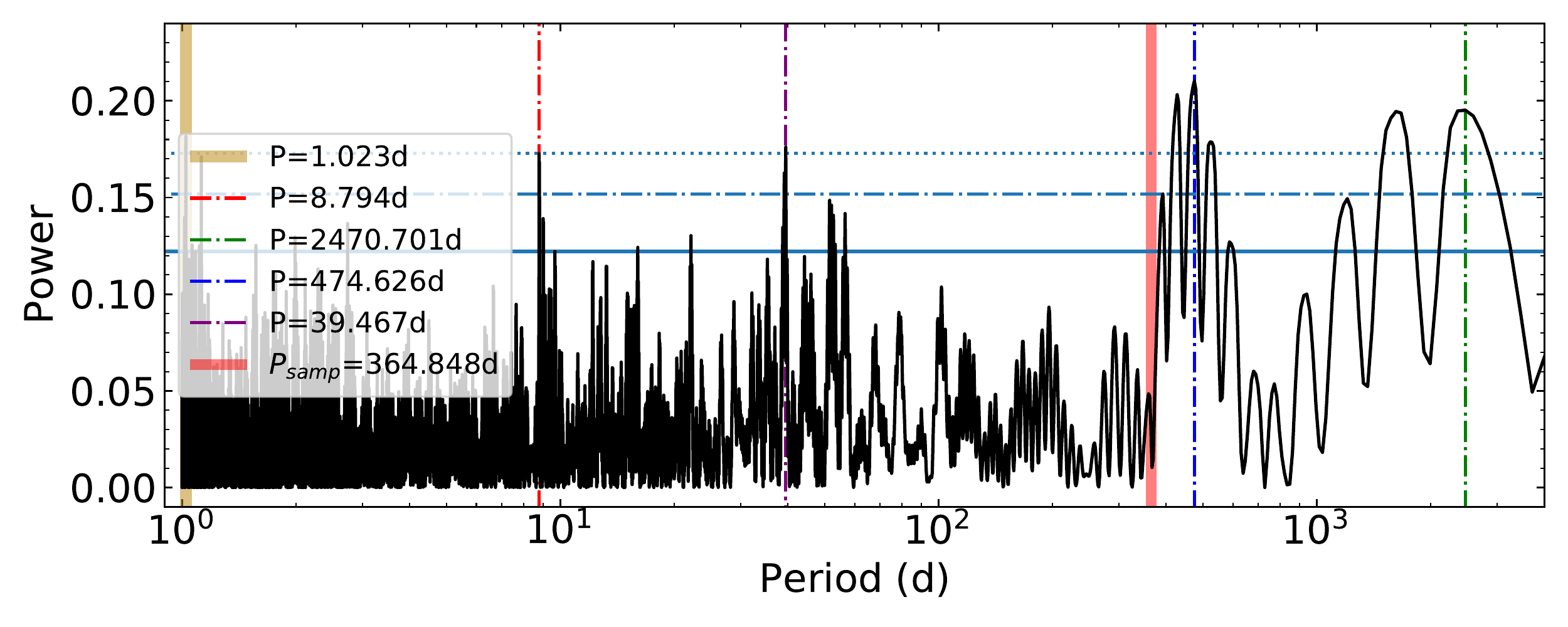}} &
\subfloat{\includegraphics[width = \columnwidth, height=3.5cm]{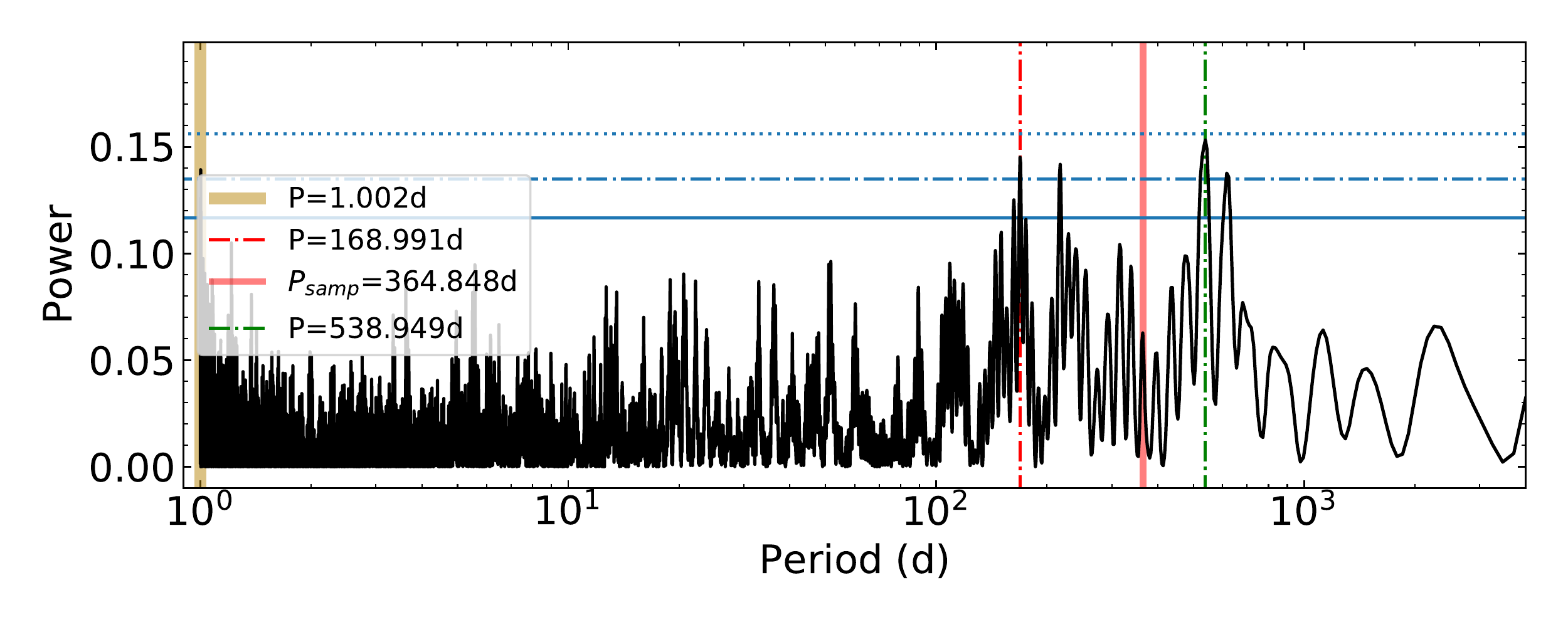}}\\
\subfloat{\includegraphics[width = \columnwidth, height=3.5cm]{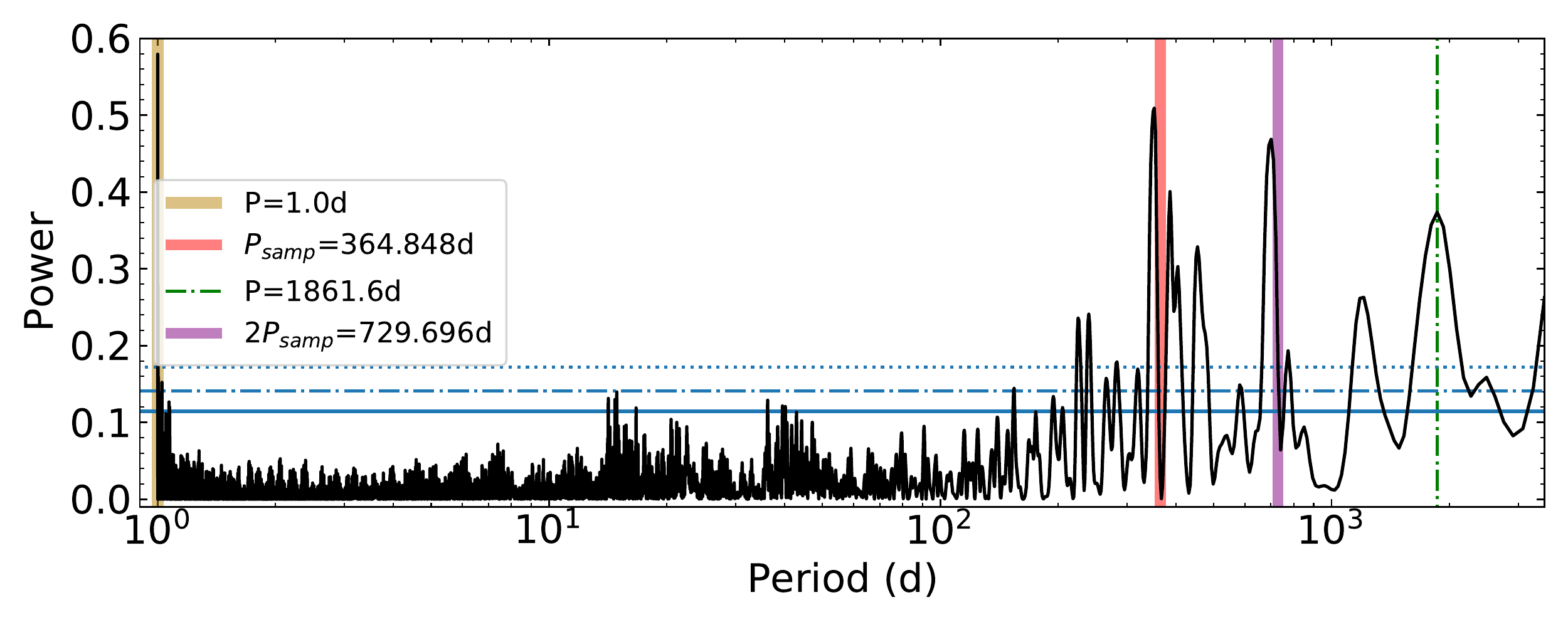}} &
\subfloat{\includegraphics[width = \columnwidth, height=3.5cm]{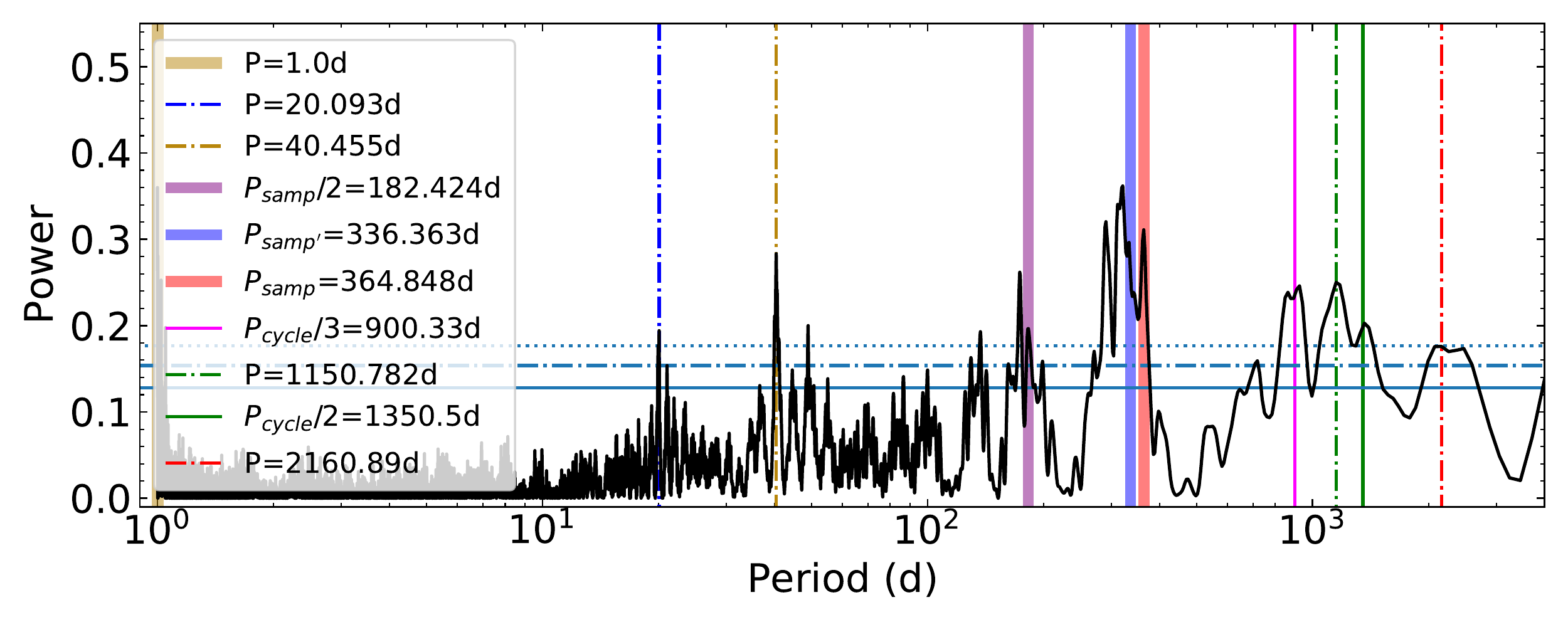}}\\

\end{tabular}
\caption{GLS periodogram of the CaII H\&K (top left), HeI (top right), NaI (bottom left) and H$\alpha$ (bottom right) indices with the x-axis showcasing the trial periods in logarithmic scale. Periods with FAP < 0.1\% are considered significant and shown using vertical coloured dash-dot lines with their respective periods mentioned in the plot legend of each periodogram. For HeI, the plot legend shows periods with FAP < 1\%. The thick coloured vertical lines show the three sampling window periods detected from the sampling window function periodogram. In the H$\alpha$ periodogram, the colored solid lines show the integer harmonics of the 7.4 years photometric activity cycle. The solid, dash-dot and dotted blue horizontal lines show the false alarm levels for 10\%, 1\% and 0.1\% respectively.}
\label{fig:GLS_all_plot}
\end{figure*}

\section{Variability of Activity indices} \label{Act_ind_periodicity_sec}
In the following, we investigate the short and long-term periodicity of activity indices in order to study the stellar rotation period as well as a possible long-term activity cycle. Recently, an activity cycle of 7.4 years was reported for GJ 436 in photometry \citep{2018AJ....155...66L}, and here we explore if it is detectable in spectroscopy. \par

\subsection{Generalised Lomb-Scargle (GLS)} \label{sec:GLS_sec}

For our analysis, we use the Generalised Lomb-Scargle (GLS) method \citep{2009A&A...496..577Z}. Effectively sampled period grid ranging from 1d to 4000d is used in order to cover at least one oscillation of the 7.4 years activity cycle. The false alarm probability (FAP) of each detected period is estimated using the 'bootstrap' method \citep{1997A&A...320..831K}. The error on each detected period is determined from the full width at half maximum (FWHM) of its respective power peak in the periodogram. Periods detected with FAPs < 0.1 \% are considered significant. \par

For HeI, NaI \& H$\alpha$, we combine the indices from the HARPS \& NARVAL datasets. The NARVAL 2016 observations fill a gap in the 14 years of HARPS observations. For CaII H\&K indices, we use the HARPS dataset alone, since this index is not calculated for NARVAL (see section \ref{sec:CaII}). Figure \ref{fig:GLS_all_plot} shows the periodograms for CaII H\&K, HeI, NaI \& H$\alpha$ indices. Periodograms per observing epoch for each of these indices are included in the supplementary material and are available online. \par

\subsubsection{CaII H\&K} \label{CaIIHK_sec}
From the periodogram of CaII H\&K indices, shown in figure \ref{fig:GLS_all_plot}, we detect a period of $474.6^{+14.8}_{-24.5}$d with the highest peak in power. The next best period is 2470.7d, roughly 6.8 years, with a relatively large FWHM. In addition to these, we detect two short periods of $39.47^{+0.11}_{-0.15}$d and $8.79^{+0.02}_{-0.01}$d. \par
We detect a strong peak around 1 day. This period is identified as a direct result of our sampling window function (see Appendix \ref{Window_function_appendix}). Along with that, we detect a period of $364.8^{+5.2}_{-6.3}$d ($P_{samp}$ from hereon). This period is possibly a result of long term observations, with data collected each year. It should be noted that we do not detect it in the CaII H\&K periodogram with a significant FAP. Both $P_{samp}$ and the 1d sampling periods are shown in figure \ref{fig:GLS_all_plot} with red and gold thick lines respectively. \par
Running the periodogram analysis on each observing epoch individually, we detect periods of $43.2^{+7.8}_{-4.6}$d and $48.6^{+12.2}_{-7.2}$d from the 2007 \& 2008 epochs respectively.
These periods are in good agreement with the $39.47^{+0.11}_{-0.15}$d period detected from the combined HARPS \& NARVAL datasets. From the 2009 epoch, we detect two periods of 87.27d, with a relatively large FWHM, and $21.9^{+1.3}_{-1.4}$d. We detect a short period of $2.82^{+0.04}_{-0.07}$d from the 2008 epoch as well .

\subsubsection{NaI}
From the periodogram of NaI indices, shown in figure \ref{fig:GLS_all_plot}, we detect a period of $351.9^{+6.0}_{-9.4}$d with the highest peak in power, after the 1d sampling window period. However, this period is 3.6 \% shorter than $P_{samp}$, shown with the red thick line, and is not considered a true period. Another significant period of $699.3^{+23.3}_{-27.1}$d is detected which is roughly twice $P_{samp}$ and hence not considered a true period. In addition to these, a period of 1861.6d, roughly 5.1 years, is detected as well. \par 
Running the periodogram analysis on each observing epoch individually, we detect a period of $46.9^{+11.9}_{-6.1}$d from one epoch only, the 2008 epoch. However, we detect shorter periods of $14.90^{+0.78}_{-0.58}$d and $14.92^{+0.82}_{-0.70}$d from the 2008 and 2016 epochs, respectively.

\subsubsection{\texorpdfstring{H$\alpha$}{HaI}} \label{HaI_periodicity_sec}
From the periodogram of H$\alpha$ indices, shown in figure \ref{fig:GLS_all_plot}, we detect a period of $321.3^{+27.0}_{-14.6}$d with the highest peak in power. But like for NaI, this period is 4.5\% shorter than the sampling window period of $336.4^{+5.7}_{-8.1}$d (see Appendix \ref{Window_function_appendix}) and, within its error bars, is most probably due to our sampling window. We detect the $P_{samp}$ period along with its $2^{nd}$ integer harmonic, $P_{samp}$/2, shown using appropriate thick coloured lines in the periodogram. In addition to these, we detect two short periods of $40.46^{+0.44}_{-0.52}$d and $20.09^{+0.04}_{-0.05}$d, with the latter being a possible $2^{nd}$ integer harmonic of 40.46d. \par
We detect the $2^{nd}$ and $3^{rd}$ integer harmonics of the 7.4 year activity cycle, shown using appropriate thick coloured lines in the periodogram, along with 2 distinct periods of 1150.8d and 2160.9d, with a relatively large FWHM. Like for CaII H\&K and NaI, the 1d sampling window period is detected as well. \par
Running the periodogram analysis on each observing epoch individually, we detect a period of $42.1^{+2.0}_{-4.3}$d and $44.1^{+8.1}_{-5.5}$d from the 2007 \& 2009 epochs respectively.
From the 2008 epoch, we detect two notable short periods of $50.3^{+8.7}_{-4.9}$d and $19.85^{+0.77}_{-0.93}$d. \par

\subsubsection{HeI} \label{HeI_periodicity_sec}

From the periodogram of HeI indices, shown in figure \ref{fig:GLS_all_plot}, we do not detect any periods with FAPs < 0.1\%. However, we do detect 2 distinct periods with FAPs < 1\%. These periods are $538.95^{+20.08}_{-27.51}$d \& $168.99^{+2.28}_{-2.03}$. We detect the sampling window period of 1d as well. Unlike the other indices, we do not detect any significant long-term periods. \par

Running the periodogram analysis on each observing epoch individually, we do not detect any periods with FAPs < 0.1\%.
\begin{figure*}
\begin{tabular}{cc}
\subfloat{\includegraphics[width = \columnwidth, height=3.5cm]{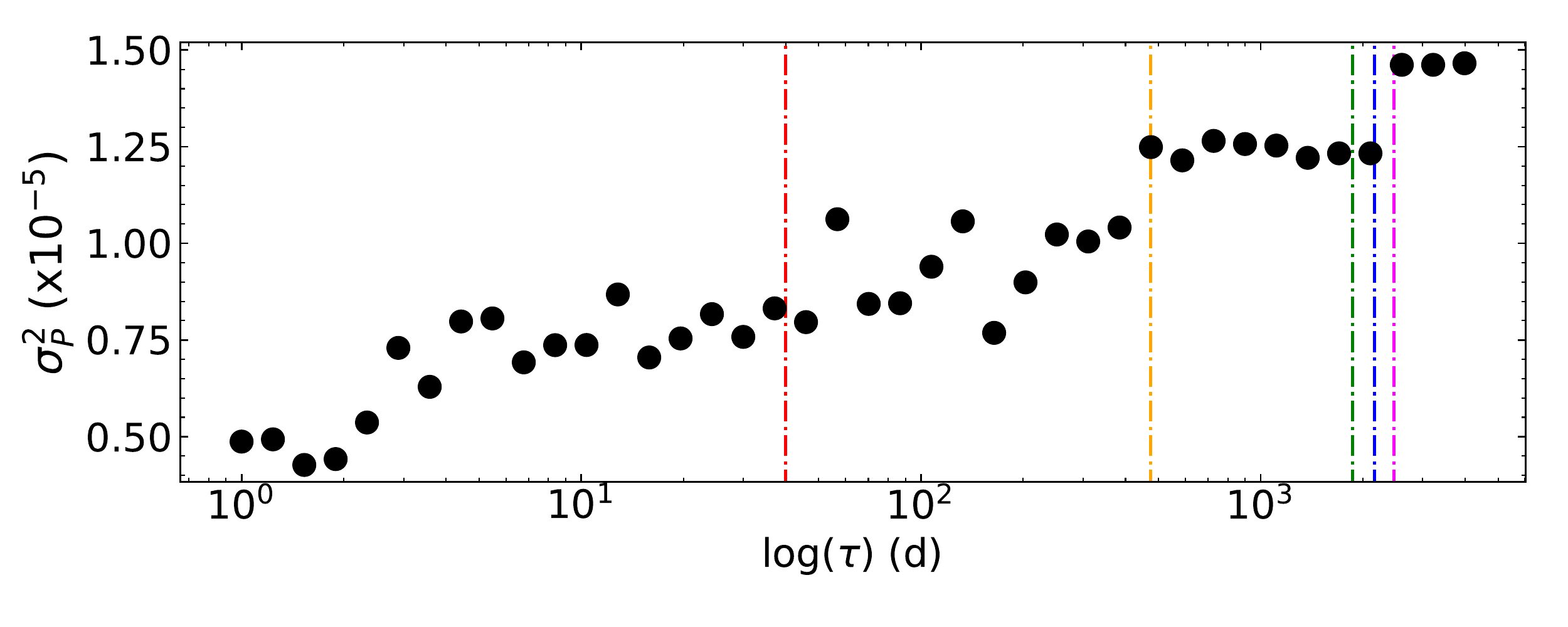}} &
\subfloat{\includegraphics[width = \columnwidth, height=3.5cm]{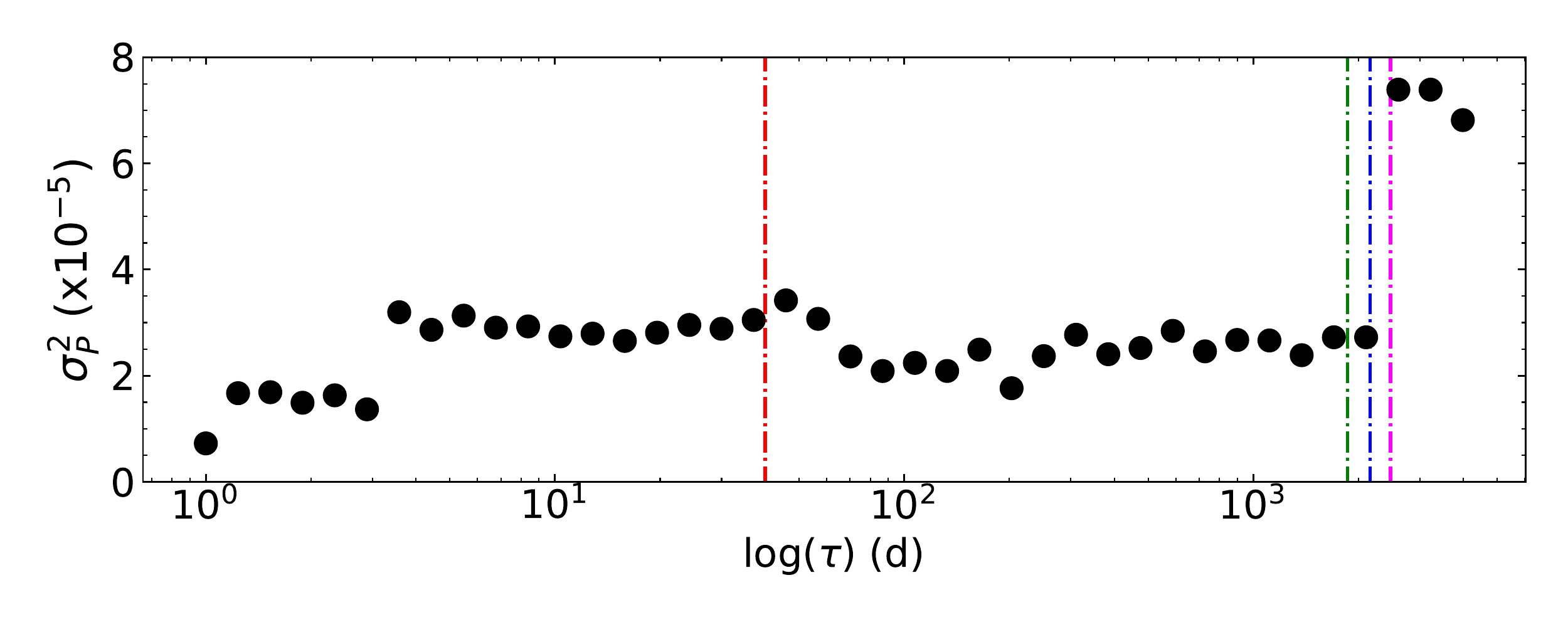}}\\

\end{tabular}
\caption{Pooled variance of H$\alpha$ (left) and NaI (right) indices calculated from both HARPS and NARVAL datasets. The x-axis shows timescales ($\tau$) in logarithmic scale ranging from 1d to 4000d. In both the figures, the red vertical line represents the stellar rotation period of 40d detected from the H$\alpha$ and CaII H\&K periodogram. The green, blue and magenta lines represent the longest periods of 1861.6d, 2160.9d and 2470.7d detected from periodograms of NaI, H$\alpha$ and CaII H\&K indices respectively. The orange vertical line in the left figure represents the 474.6d period detected from the periodogram of CaII H\&K indices (see section \ref{CaIIHK_sec})}
\label{fig:pooled_variance_plot}
\end{figure*}

\begin{figure}
    \includegraphics[width=\columnwidth]{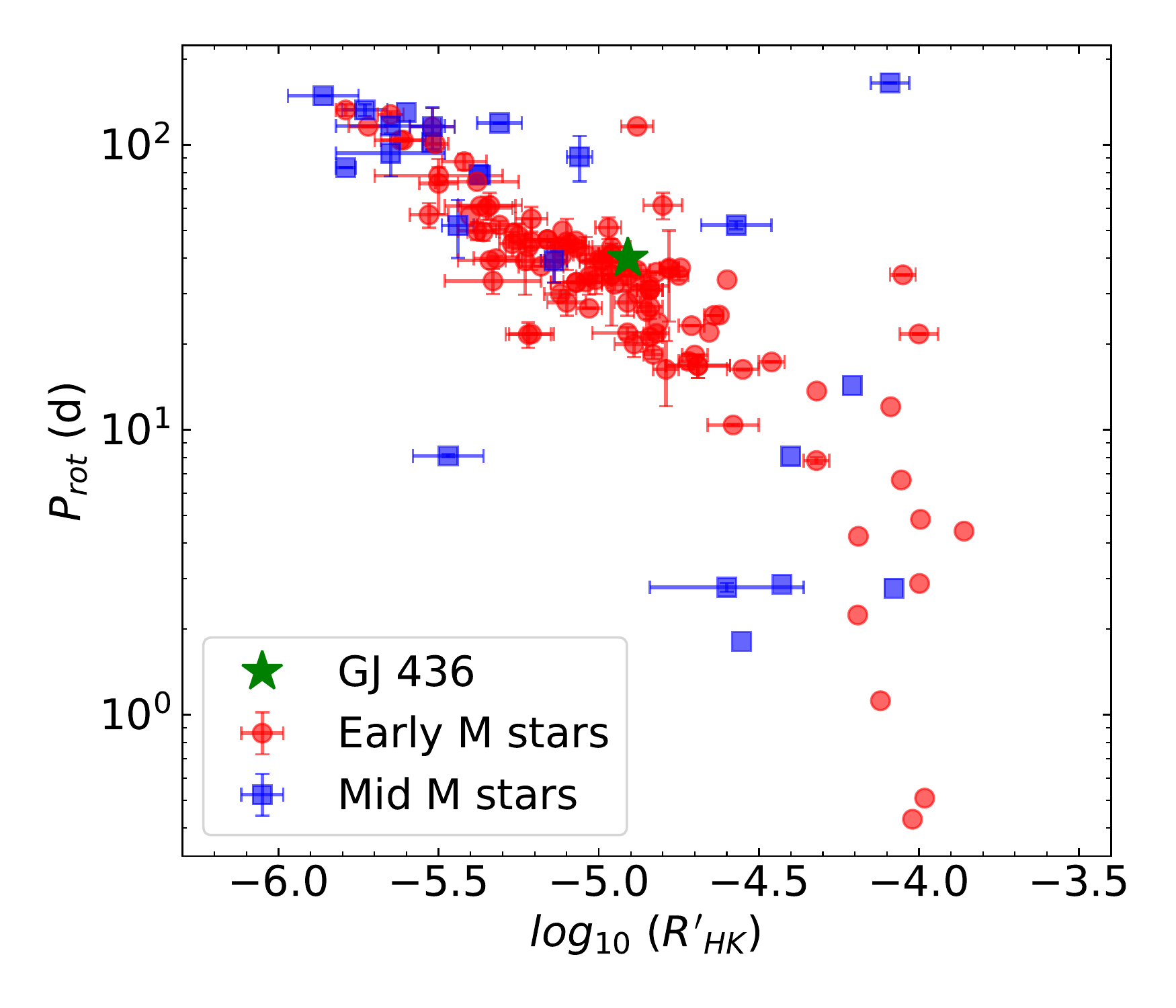}
    \caption{Chromospheric activity level $log_{10} R'_{HK}$ plotted against the stellar rotation periods of 112 M dwarfs. The red dot markers are the early M-type stars ranging from M0 - M3. The blue square markers are the mid M-type stars ranging from M3.5 to M6. The green star marker represents our star GJ 436 with its $P_{rot}$ equal to 40d and its $log_{10} R'_{HK}$ calculated as -4.909 from the CaII H\& indices. The data used to create this figure is taken from \citep{2018A&A...612A..89S, 2017A&A...600A..13A, 2017MNRAS.468.4772S, 2016A&A...595A..12S, 2015MNRAS.452.2745S}}
    \label{fig:fig_12_SM_et_al_2018}
\end{figure}

\begin{figure}
    \includegraphics[width=\columnwidth]{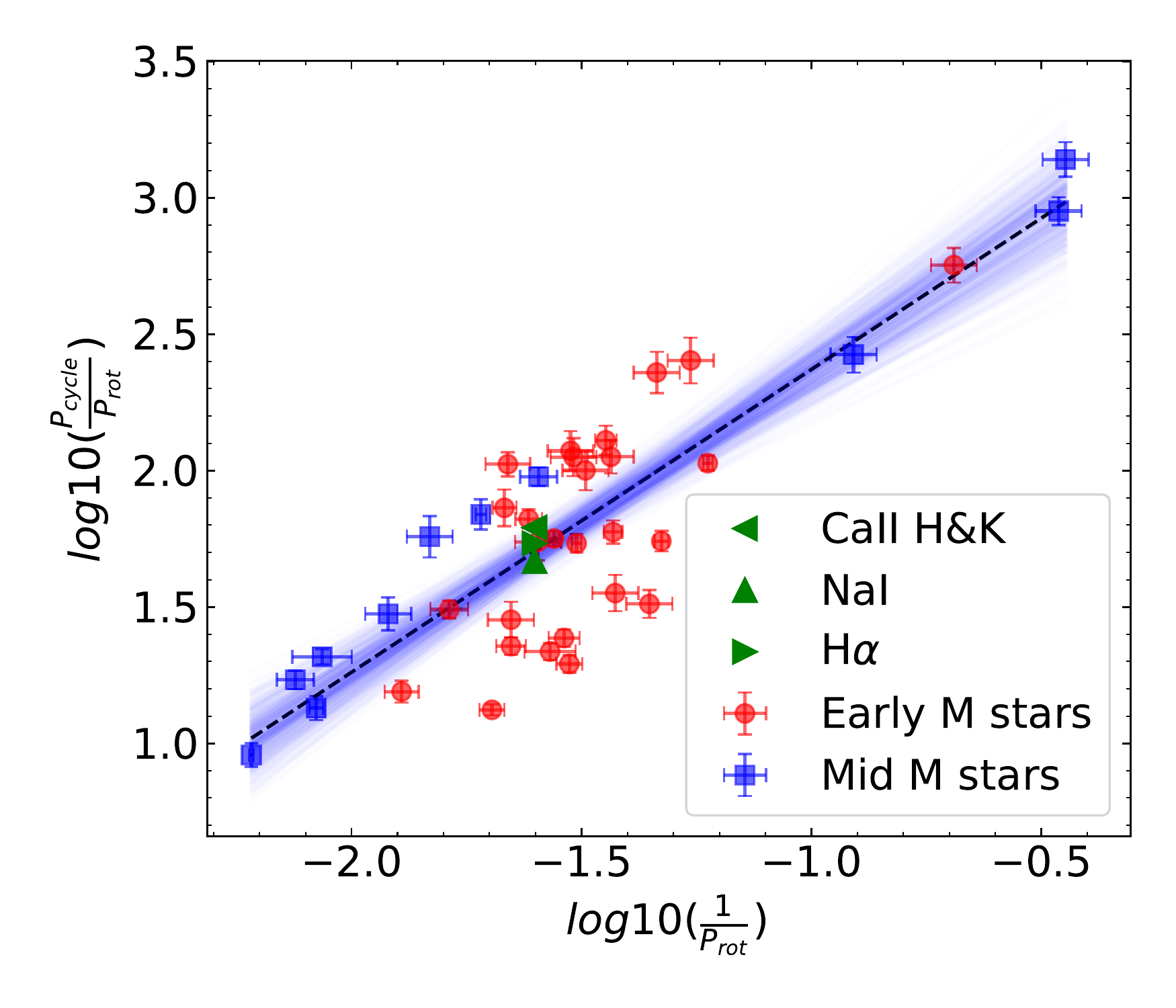}
    \caption{$log10(\frac{1}{P_{rot}})$ plotted against $log10(\frac{P_{cycle}}{P_{rot}})$ for 38 M dwarfs. The red dot markers are the early M-type stars ranging from M0 - M3 and the blue square markers are the mid M-type stars ranging from M3.5 to M6. The green triangle markers represent the long-term periods of 2470.7d, 1861.6 and 2160.9d detected from the periodograms of CaII H\&K, NaI and H$\alpha$ indices respectively. The $P_{rot}$ is equal to 40.0d in all three cases. The dashed black line represents the best-fit line and the blue lines shape a band depicting the 95\% confidence interval of the best-fit line. The data used to create this figure is taken from \citep{2018A&A...612A..89S, 2016A&A...595A..12S}}
    \label{fig:fig_13_SM_et_al_2018}
\end{figure}

\subsection{Prewhitening} \label{prewhitening_sec}

\begin{table}
    \renewcommand\thetable{6}
	\centering
	\caption{Table containing the detected periods from GLS periodograms of CaII H\&K \& H$\alpha$ indices after applying the prewhitening technique. The original short-term periods detected from both of these indices is mentioned once in red and blue bold colours. The detected periods similar to the original periods are shown using the same bold colours as well. FAPs for all periods are shown within brackets next to them.}
	\label{tab:prewhitening_results_table}
	\resizebox{\columnwidth}{!}{%
	\renewcommand{\arraystretch}{1.3}
	\begin{tabular}{ccc}
		\hline
		Prewhitened Period (d) & Detected Periods (d) & Original Period (d)\\
		\hline
		\hline
		  & \textbf{CaII H\&K} & \\
	    \hline
		474.63 (< 0.1\%) & $9.68^{+0.03}_{-0.01}$ (< 1\%) & \textcolor{red}{$\mathbf{39.47^{+0.11}_{-0.15}}$} (< 0.1\%)\\
		  & $32.36^{+0.33}_{-0.3}$ (< 1\%) & \textcolor{blue}{$\mathbf{8.79^{+0.02}_{-0.01}}$} (< 0.1\%)\\
		 & \textcolor{blue}{\textbf{8.80 ± 0.03}} (< 1\%) & \\
		\hline
		2470.70 (< 0.1\%) & $9.68^{+0.03}_{-0.01}$ (< 0.1\%) & \\
		 & \textcolor{blue}{\textbf{8.80 ± 0.03}} ($\approx$ 0.1\%) & \\
		 & $102.80^{+3.04}_{-4.80}$ (< 1\%) & \\
		 & $80.00^{+1.60}_{-2.11}$ (< 1\%) & \\
		 & \textcolor{red}{$\mathbf{38.81^{+1.40}_{-0.64}}$} (< 10\%) & \\
	    \hline
		  & \textbf{H$\alpha$} & \\
		\hline  
		1150.78 (< 0.1\%) & $137.12^{+1.33}_{-1.2}$ (<0.1\%) & \textcolor{red}{$\mathbf{40.46^{+0.44}_{-0.52}}$} (< 0.1\%)\\
		 & $48.96^{+0.17}_{-0.23}$ (< 0.1\%) & \textcolor{blue}{$\mathbf{20.09^{+0.04}_{-0.05}}$} (< 0.1\%)\\
		 & $173.53^{+1.93}_{-1.91}$ (< 0.1\%) & \\
		 & $55.06^{+0.22}_{-0.19}$ (< 0.1\%) & \\
		 & \textcolor{blue}{$\mathbf{19.99^{+0.12}_{-0.11}}$} (< 0.1\%) & \\
		 & \textcolor{red}{$\mathbf{40.46^{+0.23}_{-0.66}}$} ($\approx$ 1\%) & \\
		\hline 
		2160.89 (< 0.1\%) & $137.00^{+1.46}_{-1.35}$ (< 0.1\%) & \\
		 & $189.10^{+2.63}_{-4.70}$ (< 0.1\%) & \\
		 & \textcolor{blue}{$\mathbf{20.0^{+0.13}_{-0.11}}$} (< 0.1\%) & \\
		 & \textcolor{red}{$\mathbf{40.43^{+0.15}_{-0.80}}$} (< 0.1\%) & \\
		\hline
	\end{tabular}%
	}
\end{table}

With a focus on inspecting whether the short-term periods detected from the periodograms of CaII H\&K and H$\alpha$ indices, in section \ref{sec:GLS_sec}, are detectable in the absence of their respective long-term periodicity, we perform a prewhitening procedure on our indices. Following \cite{2011A&A...533A...4B}, we fit a sine wave function to the index time series, with a period equal to the long-term period. The fitted sine function is then subtracted from the index time series and a GLS periodogram is run on this "prewhitened" data to reassess its short-term periodicity. When multiple long-term periods are detected in a periodogram, we apply the prewhitening procedure on each of these periods separately. \par

For the CaII H\&K indices, two long-term periods of 474.63d and 2470.7d are detected in the GLS periodogram. When we prewhiten the 474.63d period, we detect a short-term period of 32.36d, which is $\approx$ 18\% shorter than the original period of 39.47d. When prewhitening the 2470.7d period, we detect a period of $38.81^{+1.40}_{-0.64}$d. This period, within its error bars, is comparable to the 39.47d period detected from the non-prewhitened CaII H\&K indices but with a larger FAP of < 10\%. \par

For the H$\alpha$ indices, we prewhiten the 1150.78d period and from the resultant periodogram, we detect a period of $40.46^{+0.23}_{-0.66}$d. This period is exactly the same as the original 40.46d period detected from the non-prewhitened H$\alpha$ indices but has a different FAP of $\approx$ 1\% instead of < 0.1\%. When we prewhiten the long-term period of 2160.89d, we detect a period of $40.43^{+0.15}_{-0.80}$d. This period, within its errors bars, is comparable to the 40.46d period with both having the same FAP of < 0.1\%. \par

Table \ref{tab:prewhitening_results_table} shows all of the detected periods after prewhitening for both CaII H\&K and H$\alpha$ indices with detected periods similar to the original period shown in red and blue bold colours.

\subsection{Pooled Variance Analysis}

In order to further investigate the long-term periodicity of our activity indices, we implemented the "Pooled Variance" (PV) technique \citep{2004A&A...425..707L, 1997SoPh..171..191D, 1997SoPh..171..211D}. The PV is calculated by binning our data into evenly-spaced bins with some timescale length $\tau$. The mean of the variance from each bin for a given $\tau$ is the PV for that $\tau$. Changes in the PV for different timescales are assumed to be associated with several mechanisms such as stellar rotation, the lifetime of possible Active Region (AR), and long-term activity cycles. The PV technique can thus be used to identify the approximate timescales at which a given activity mechanism dominates. 
We apply PV to NaI \& H$\alpha$ indices. We do not apply it to CaII H\&K indices, because CaII H\&K indices were not calculated for NARVAL (see section \ref{CaII_H_sec}), and thus the gap in our data is of 10 years. HeI does not show any significant long-term periodicity for us to investigate it further. Figure \ref{fig:pooled_variance_plot} shows the PV diagrams for H$\alpha$ (left) and NaI (right) indices for timescales ranging from 1d to 4000d. \par

For H$\alpha$ indices, the PV diagram at shorter timescales increases steadily. Near the stellar rotation timescale of 40d, shown using the red vertical line in figure \ref{fig:pooled_variance_plot}, the PV appears to be relatively constant over a short timescale range. Eventually, the PV reaches a plateau starting at $\approx$ 470d which could be related to the lifetime of potential AR complexes for GJ 436. This period fits well within the range of AR lifetimes observed for slow M rotators \citep{2017MNRAS.472.1618G}. This timescale is comparable to the period of 474.6d as well, detected from the periodogram of CaII H\&K indices (see section \ref{CaIIHK_sec}), shown using the orange vertical line. At timescales greater than 1000d, we observe a sharp increase in the PV reaching another plateau starting at $\approx$ 2500d. This PV increase is likely due to a possible activity cycle. To compare it with the long-term periods detected from section \ref{Act_ind_periodicity_sec}, we show the periods of 1861.6d, 2160.9d \& 2470.7d, detected from NaI, H$\alpha$ \& CaII H\&K periodograms respectively, using appropriately coloured vertical lines in figure \ref{fig:pooled_variance_plot}. The 2470.7d period detected from the CaII H\&K periodogram, being close to the period of 2160.9d detected from the H$\alpha$ periodogram, falls near the start of the activity cycle plateau and is in good agreement with the $\approx$ 2500d period.

For NaI indices, the PV diagram remains constant for timescales up to 3d and then nearly doubles. With increasing timescales, the PV remains constant until a timescale close to the stellar rotation where it slightly decreases to eventually have a sharp peak at the timescale of $\approx$ 2500d. \par

\section{Discussion} \label{Discussion_sec}

In this paper, we studied the chromospheric activity of GJ 436 using data collected over 14 years by the HARPS spectrograph and the NARVAL spectropolarimeter (spanning 85 days), with the NARVAL data falling within the 14 years of HARPS observations. We analysed six different spectral lines: CaII H\&K, HeI, NaI, H$\alpha$ CaI and CaII IRT, and the correlations between them. We explored the periodicity of these indices, both on a short and long time-scale. \par 

\subsection{Index Correlations}

From the long time span HARPS dataset, we found the CaII H\&K indices to show a weak positive correlation with the H$\alpha$ indices. The HeI index was found to show a weak positive correlation with the H$\alpha$ index as well. In order to explore how these correlations change over time, they were studied for each of the six observing epochs of HARPS , as well as the 2016 epoch of NARVAL. 

For two pairs of correlations, we found at most two observing epochs which showed significant correlations, i.e. with p-values less than 0.05. The coefficient of correlation for these pairs seem to be decreasing with the mean activity level of the star (calculated for one of the activity indices), shown in figure \ref{fig:corr_per_epoch_plots}. However, we have only 2 significant correlations for each of these correlation pairs, and caution the reader that this trend needs further confirmation. The differences in correlation between epochs could be due to different phases of an activity cycle \citep[see, e.g.][]{2017A&A...598A..28S, 2009A&A...501.1103M}. In fact, GJ 436 has a reported photometric activity cycle of 7.4 years \citep{2018AJ....155...66L}. \cite{2011A&A...534A..30G}, when studying long term activity of a sample of M dwarfs, found the correlations to vary with the level of stellar activity. Change of correlation between indices is reported for the Sun as well \citep{2019A&A...627A.118M}, with these changes being attributed to the variation of the concentration of plages and dark filaments on the solar surface. \par

From the H$\alpha$ vs CaII plot in figure \ref{fig:corr_per_epoch_plots}, we found a strong positive correlation of 0.9 at the lowest mean activity level of the CaII indices. The CaII lines are more sensitive to the surface coverage of plages than the H$\alpha$ line, which is sensitive to dark filaments \citep{2009A&A...501.1103M, 2011A&A...534A..30G}. The existence of plages and dark filaments in the same location on the stellar surface could thus account for the strong positive correlation, occurring in the 2010 observing epoch. \par
From the 2016 NARVAL epoch, we found the CaII IRT$_{3}$ to correlate negatively with the CaII IRT$_{2}$. The IRT$_{3}$ showed a negative correlation with the H$\alpha$ index as well. \par

\subsection{Stellar Rotation Period}

Using the combined HARPS and NARVAL dataset, we detected short-term periods of $39.47^{+0.12}_{-0.15}$d and $40.46^{+0.44}_{-0.52}$d from the variations of CaII H\&K and H$\alpha$ activity indices respectively. Both periods have FAPs < 0.1 \% and are considered significant. These periods are interpreted as the rotation periods of GJ 436  and are in good agreement with the 39.9 ± 0.8d period found by \cite{2015MNRAS.452.2745S} in the variations of $log_{10}(R'_{HK})$ and the 42.6 ± 2.2d period found by \cite{2018Natur.553..477B} in the variations of H$\alpha$ indices. Exploring periodicity in each of the 7 observing epochs individually, we detected periods of $43.2^{+7.8}_{-4.6}$d \& $48.6^{+12.2}_{-7.2}$d from CaII H\&K and $42.1^{+2.0}_{-4.3}$d \& $44.1^{+8.1}_{-5.5}$d from H$\alpha$ indices in 2 separate observing epochs. These periods, within their error bars, are compatible with the ones found from the whole dataset. To further test the effect of long-term periodicity on the detection of these rotation periods, we prewhitened the long-term periods (see section \ref{prewhitening_sec}) for the CaII H\&K and H$\alpha$ indices. We detected the rotation periods in the prewhitened data as well, with varying FAPs, showcasing that these periods are thus significant and not aliases of the long-term periodicity. \par

Considering the B - V color of GJ 436, these derived $P_{rot}$ periods fall within the expected range of detected rotation periods for early M dwarfs of spectral type M0-M3 \citep[see figure 10 in ][]{2018A&A...612A..89S}. These periods are consistent with the rotation periods of M dwarfs with stellar masses similar to GJ 436 as well \citep[see figure 10 in][]{2017ApJ...834...85N}. From this figure, our star falls within the inactive M dwarfs branch. To compare our mean $P_{rot}$ of 40.0d from the variations of CaII H\&K and H$\alpha$ indices to other M dwarfs, we calculated a mean $log_{10} (R'_{HK})$ for GJ 436 and plot it against the stellar rotation periods of 112 M dwarfs \citep{2018A&A...612A..89S, 2017A&A...600A..13A, 2017MNRAS.468.4772S, 2016A&A...595A..12S, 2015MNRAS.452.2745S}, ranging in spectral type from M0 - M3 (see fig.\ref{fig:fig_12_SM_et_al_2018}). GJ 436 falls within the trend of the M dwarfs of similar rotation periods and activity levels, within the unsaturated regime of the rotation-activity relation. \par

From the NaI index variations we detect a period of $46.9^{+11.9}_{-6.1}$d, from the 2008 observing epoch, compatible with the ones detected from CaII H\&k and H$\alpha$ indices. From the 2016 observing epoch, we detect a period of $14.92^{+1.23}_{-0.98}$d comparable to $\approx$ $\frac{1}{3^{rd}}$ of $46.9^{+11.9}_{-6.1}$d, within its error bars. \par

\subsection{Activity Cycle Period}

The CaII H\&K, NaI \& H$\alpha$ indices show significant long-term periodicity, which one might expect given the reported photometric activity cycle of 7.4 years \citep{2018AJ....155...66L}. The indices show long-term variations with periods of 2470.7d, 1861.6d \& 2160.9d, which are roughly 6.8 years, 5.1 years and 5.9 years respectively. Additionally, we detect a period of $474.6^{+14.8}_{-24.5}$d from the CaII H\&K index variations. From the H$\alpha$ index variations, we detect the $2^{nd}$ and $3^{rd}$ integer harmonics of the photometric activity cycle as well with periods 1350.5d and 900.3d respectively. \par
We reassessed the long-term periodicity of H$\alpha$ and NaI indices  using the "Pooled Variance" technique which indicates the lifetime of an activity cycle to be $\approx$ 2500d. This is in good agreement with the long-term period of 2470.7d detected from the CaII H\&K periodogram. We compare our derived cycle periods to other M stars to place them in the perspective of early M dwarfs. Using 40.0d as our derived $P_{rot}$, we compare $log_{10}$ of $\frac{P_{cycle}}{P_{rot}}$ against $log_{10}$ of $\frac{1}{P_{rot}}$ for 38 M dwarfs. Here, $P_{cycle}$ are the three long-term periods derived from the CaII H\&K, NaI \& H$\alpha$ index variations. As shown in figure \ref{fig:fig_13_SM_et_al_2018} with appropriate green triangle markers, our derived $P_{cycle}$ periods fall within the 95\% confidence interval of the best-fit line. Considering the color in this case, our derived $P_{cycle}$ periods show similar values to those of M0 - M3 type M dwarfs \citep[see figure 11 in][]{2018A&A...612A..89S}.

\section{Conclusion} \label{Conclusion_sec}

From our results, we detect the stellar rotation period of GJ 436 from the variations of spectroscopic activity indicators. \cite{2018A&A...612A..89S} find a mean stellar rotation \& activity cycle period of 33 ± 23 d \& 6.3 ± 3.4 years respectively, for M0 - M3 type M dwarfs, with both our $P_{rot}$ and $P_{cycle}$ periods falling within these expected ranges. Additionally, we find that the correlations between activity indices vary with the level of stellar activity over time. \par
The activity of GJ 436 could be following a long-term cycle, compatible with the photometric one. However, in order to better determine the cycle period, further data with denser spectroscopic monitoring are needed. Understanding magnetic activity and cycles in M-dwarfs is a crucial step in filtering the activity RV signature, which would help confirm additional planets in a system, if present. Furthermore, a better understanding of chromospheric activity cycles of GJ 436 is necessary in studying the impact of its stellar activity on the atmosphere and potential habitability of its hot-Neptune GJ 436b, which is one of the prime targets for multiple James Webb Space Telescope Guaranteed Time Observations (GTO) programs as well as other space missions such as ARIEL \citep{2022arXiv220505073E}. 

\section*{Acknowledgements} \label{Acknowledgements_sec}
We thank the anonymous referee for the constructive and valuable comments that helped improve our manuscript. We acknowledge support from the United Arab Emirates University (UAEU) startup grant number \textbf{G00003269}. We thank João Gomes da Silva for insightful discussions about the $\tt ACTIN$ python package and Julien Morin, Gaetano Scandariato, Sudeshna Boro Saikia, Claire Moutou and Andrew Cameron for various discussions about the work. This work made use of the following software: $\tt numpy$ \citep{harris2020array}, $\tt astropy$ \citep{astropy:2018}, $\tt matplotlib$ \citep{Hunter:2007}, $\tt pandas$ \citep{mckinney-proc-scipy-2010}, $\tt PyAstronomy$ \citep{pya} and $\tt scipy$ \citep{2020SciPy-NMeth}.

\section*{Data Availability}

This paper uses data based on observations collected at the European Southern Observatory under ESO programs \textbf{072.C-0488(E), 082.C-0718(B), 183.C-0437(A), 1102.C-0339(A)} and \textbf{1102.C-0339(F)}, using the HARPS spectrograph. Additionally, it uses data from the Pic du Midi Observatory, for the program \textbf{L161N04}, using the NARVAL spectropolarimeter. 

The Python code used to calculate the activity indices is publicly available on \url{https://github.com/MXK606/krome}. The data used in this paper are publicly available on ESO archive (\url{http://archive.eso.org/scienceportal/home}) for HARPS data, and on Polarbase database (\url{http://polarbase.irap.omp.eu}) for NARVAL data with additional figures available as online supplementary material.


\bibliographystyle{mnras}
\bibliography{bib_list} 



\appendix

\section{\texorpdfstring{H$\alpha$}{HaI} index method comparison}
From the NARVAL dataset for comparison, in addition to the standard method in section \ref{H_alpha_index_sec}, we adopted the method of \cite{2011A&A...534A..30G} for calculating the H$\alpha$ index where instead of mean in equation \ref{H_alpha_index_eq}, each flux was calculated as a sum over their respective bandwidths. Additionally, following \cite{2018A&A...614A..76J}, we calculated the pseudo-Equivalent Width (pEW) of the H$\alpha$ absorption line as well to compare with the H$\alpha$ index as 

\begin{equation} \label{pEW_eq}
pEW = \int_{\lambda_{1}}^{\lambda_{2}} (1 - \frac{F(\lambda)}{F_{pc}}) \,d\lambda\ 
\end{equation}
where $F_{pc}$ is defined as the mean of the median flux in the so called pseudo-continuum within the spectral ranges (6545.0Å - 6559.0Å) and (6567.0Å - 6580.0Å). For our star, the pEW value was measured within a narrower 1.6Å wide bandwidth from (6562.008Å - 6563.608Å) centered on the H$\alpha$ line than the one used by \cite{2018A&A...614A..76J}.  The error on $pEW$ was calculated following \cite{2018A&A...614A..76J}. \par
The H$\alpha$ activity index in literature is calculated simply using equation \ref{H_alpha_index_eq}. But, each flux in this equation is calculated differently by different authors. \cite{2009A&A...495..959B} calculated each flux as the mean flux value within the respective bandwidth for its target HD 189733. Similarly, \cite{2016A&A...594A..29B} calculated the flux as the mean flux value as well for its target HD 201091. \cite{2013ApJ...764....3R} however calculated each flux as the sum of the flux values for its (K5-M5) dwarf sample and \cite{2011A&A...534A..30G} did the same for its (M0-M5.5) dwarf sample. \par
\begin{figure}
\begin{tabular}{c}
\subfloat{\includegraphics[width = 0.85\columnwidth]{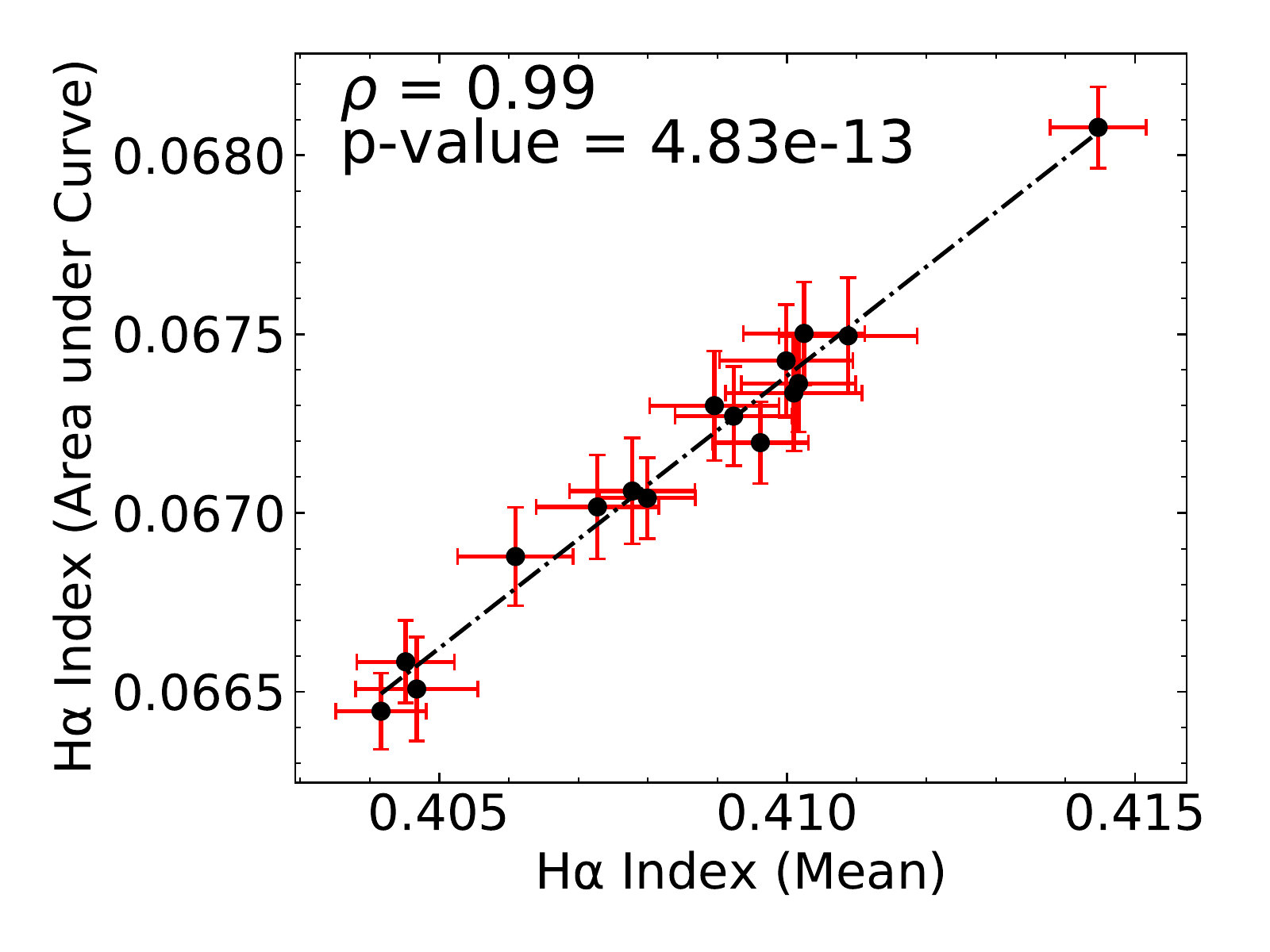}}\\ 
\subfloat{\includegraphics[width = 0.85\columnwidth]{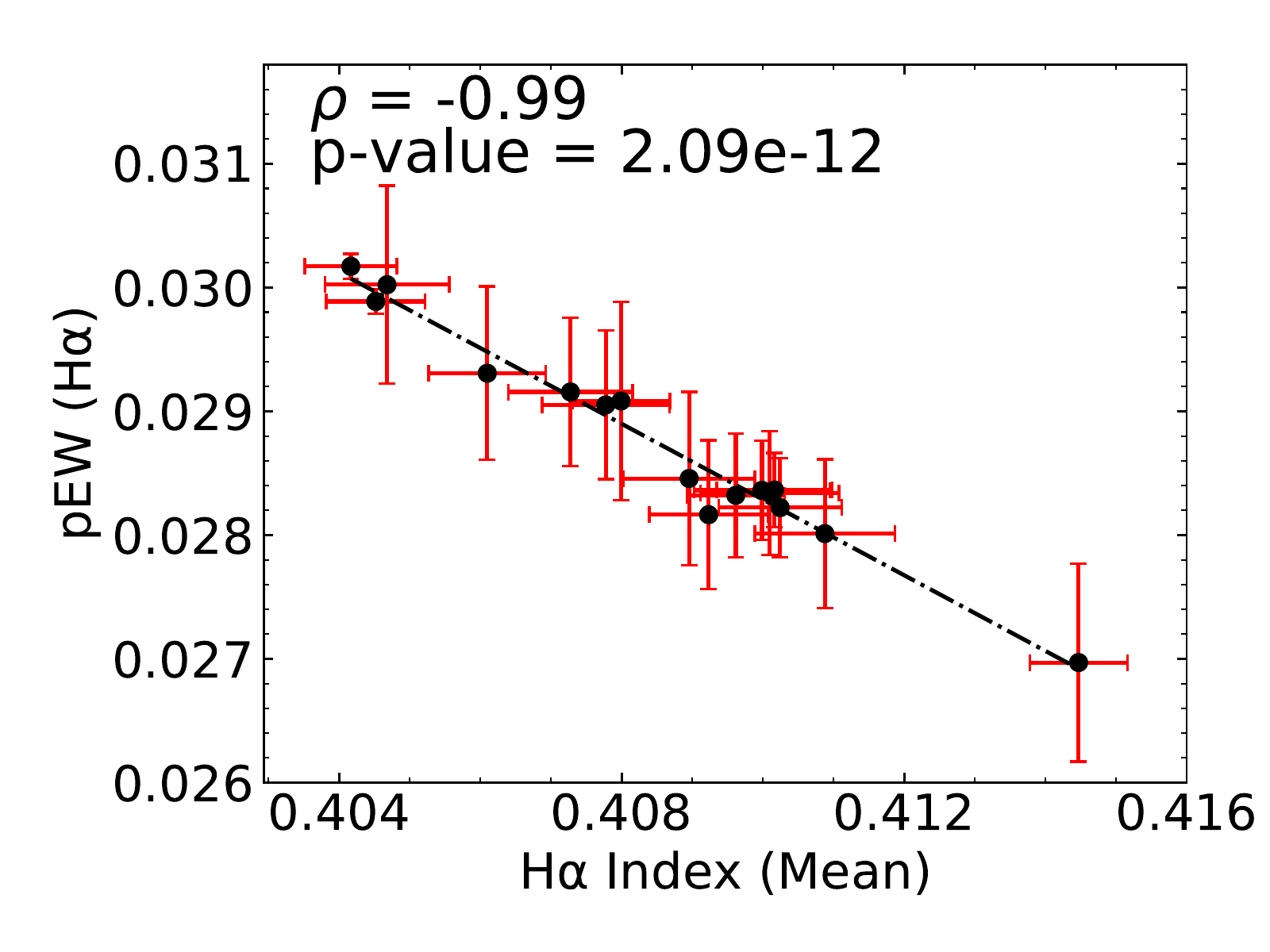}}\\
\end{tabular}
\caption{Correlation between the H$\alpha$ index calculated from the NARVAL dataset using the mean and the sum method (top) \& between H$\alpha$ index (mean method) and the pEW (bottom). The dash-dotted line shows the best-fit line and $\rho$ is the Pearson R correlation coefficient. The corresponding error bars for each index are plotted in red.}
\label{fig:H_alpha_method_comparison_plot}
\end{figure}

To check how these methods vary with one another, we compared the mean and the sum method of calculating the H$\alpha$ index by calculating its Pearson R correlation coefficient. We found their R coefficient to be 0.99 with a p-value of $5 \times 10^{-13}$, which is the probability of finding the same coefficient for a null hypothesis. Both of these index methods were found to correlate positively as shown in figure \ref{fig:H_alpha_method_comparison_plot} with the mean method producing indices $\approx$ 6 times larger than the sum method. This is simply because the sum method of index calculation is sensitive to the number of flux points in each bandwidth as opposed to the mean method and the H$\alpha$ line core bandwidth is much smaller than the reference continuum bandwidths leading to a relatively smaller sum index value. But for studying stellar activity variations, either of these methods could be employed since they both show a strong positive linear correlation. \par
The H$\alpha$ index (mean method) is then compared to the pEW and their Pearson R correlation coefficient is found to be -0.99 (p-value=$2 \times 10^{-12}$). These two methods are found to correlate negatively as shown in figure \ref{fig:H_alpha_method_comparison_plot}. The H$\alpha$ index and the pEW methods are both calculated within the same bandwidth of 1.6Å. centered on the H$\alpha$ line apart from the continuum reference bands. This shows that as the mean flux within the H$\alpha$ line increases thereby increasing the H$\alpha$ index, its relative pEW decreases, as expected. \par 

\section{NARVAL spectral orders around Calcium II H\&K} \label{CaII HK_appendix_sec}

\begin{figure}
\begin{tabular}{c}
\subfloat{\includegraphics[width = 0.98\columnwidth, height=3.5cm]{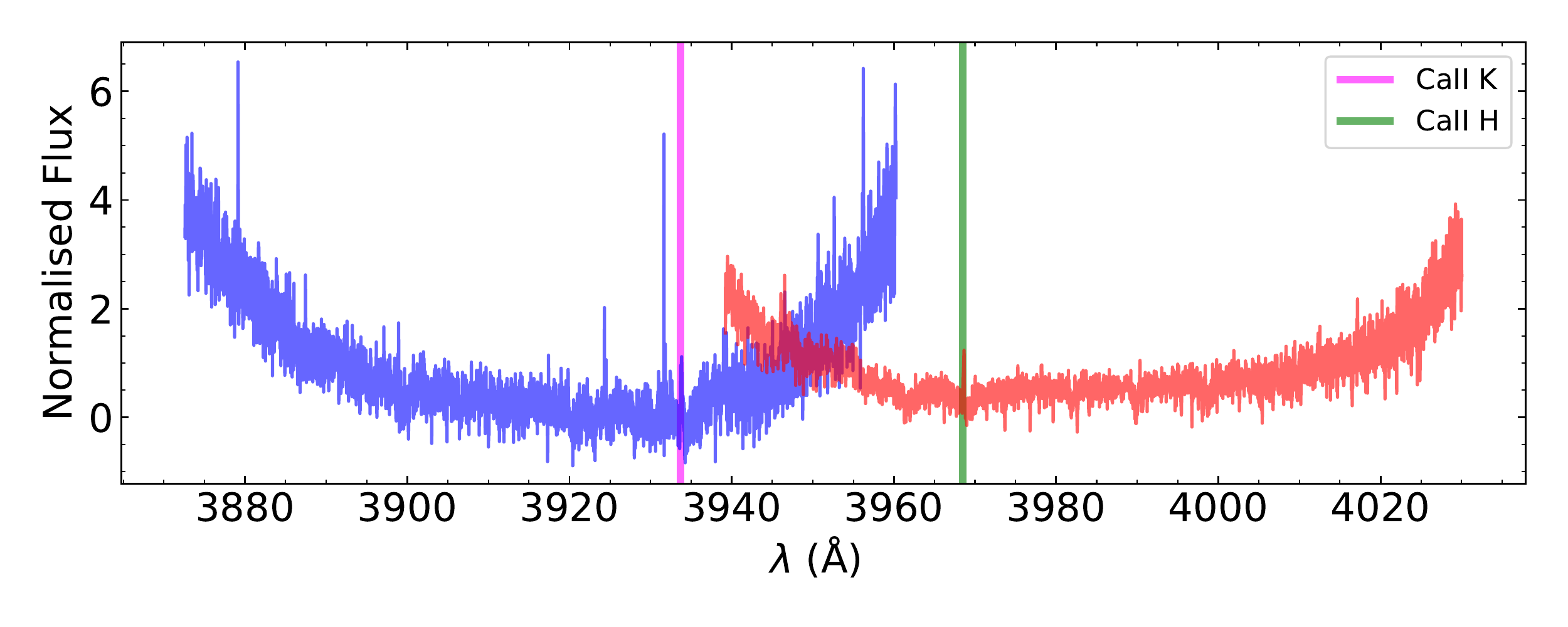}}\\
\subfloat{\includegraphics[width = 0.98\columnwidth, height=3.5cm]{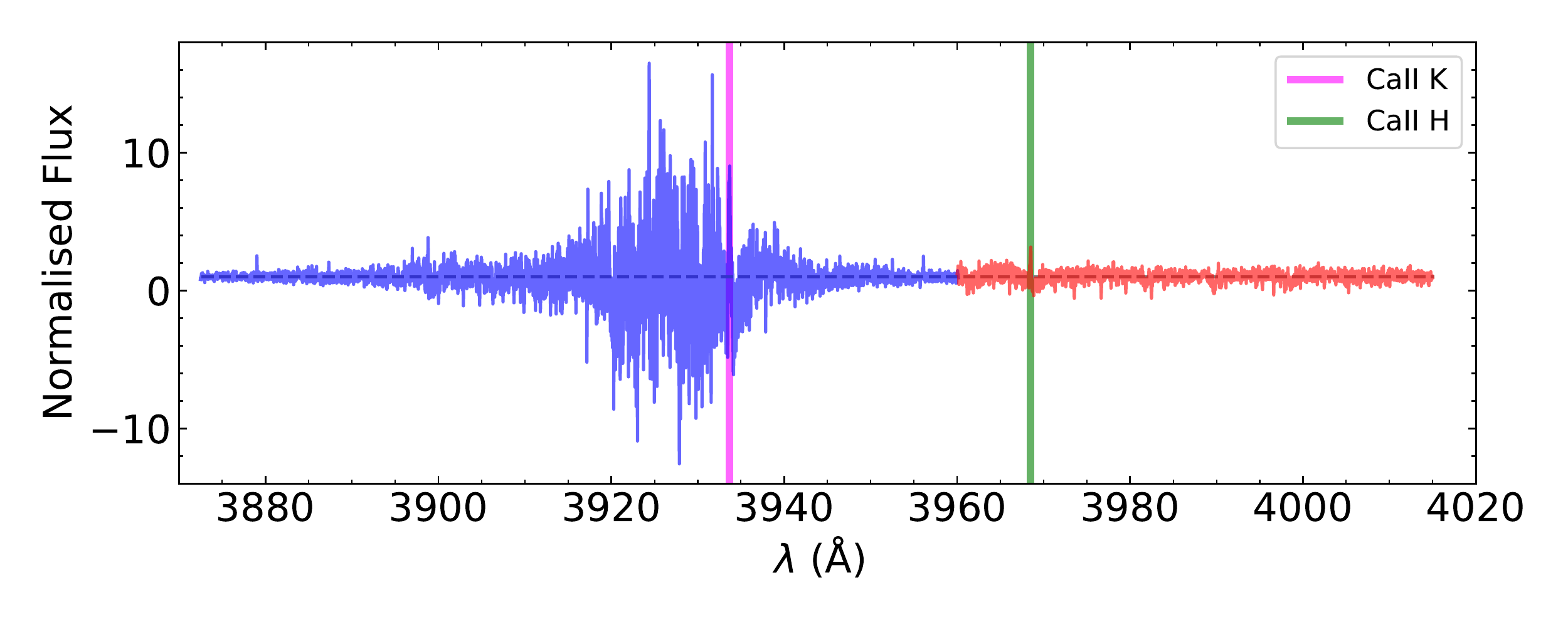}}\\
\end{tabular}
\caption{Pipeline normalised spectra (top) and the re-normalised spectra (bottom) for the two spectral orders around the CaII H\&K emission lines of the 16th March 2016 NARVAL observation. The blue spectral order contains the CaII K line and the red spectral order contains the CaII H line. The gray dashed line in the bottom plot shows the flux level at unity.}
\label{fig:pipeline_vs_renormalised_orders_CaII_plot}
\end{figure}

The NARVAL data reduced by the data reduction software LIBRE-ESpRIT contains 40 individual spectral orders numbered from \#22 to \#61. These orders overlap with one another at their respective borders, as shown in figure \ref{fig:pipeline_vs_renormalised_orders_CaII_plot}. The figure shows the spectral order containing the CaII K line to be noisier than the order containing the CaII H line. In order to calculate the CaII H\&K activity index for our star, these spectral orders need to be re-normalised to unit flux. Each spectral order was fitted to a 4th order polynomial using the astropy package $\tt specutils$ \citep{2021zndo...4603801E}. The order was then divided by this polynomial fit to give the re-normalised spectra. For the blue order containing the CaII K line, 2 strong emission lines at 3931.6Å and 3924.2Å were removed by fitting a Gaussian to them and subtracting the Gaussian fit from the entire spectra. The removal of these lines did not affect the re-normalisation process and the order containing the CaII K line was not well normalised. The reason this order looks this way after the re-normalisation procedure is due to large number of negative flux values encountered in the wavelength range of 3900 - 3950 Å. Thus when this region of the order is divided by a constant polynomial fit, which for this region has values close to zero, it stretches the spectra vertically in such fashion. Hence, only the CaII H line was analysed in section \ref{CaII H_index_eq}. 

\section{Computing the window function to check for sampling biases in periodogram} \label{Window_function_appendix}
For unevenly sampled data, often times the true period peak is surpassed in power by spurious period peaks. Such peaks can be seen in the periodograms in figure \ref{fig:GLS_all_plot}. One major factor leading to these peaks is the duration between the observations \citep{2018ApJS..236...16V}. \par
\begin{figure}
	\includegraphics[width = 0.98\columnwidth, height=4.0cm]{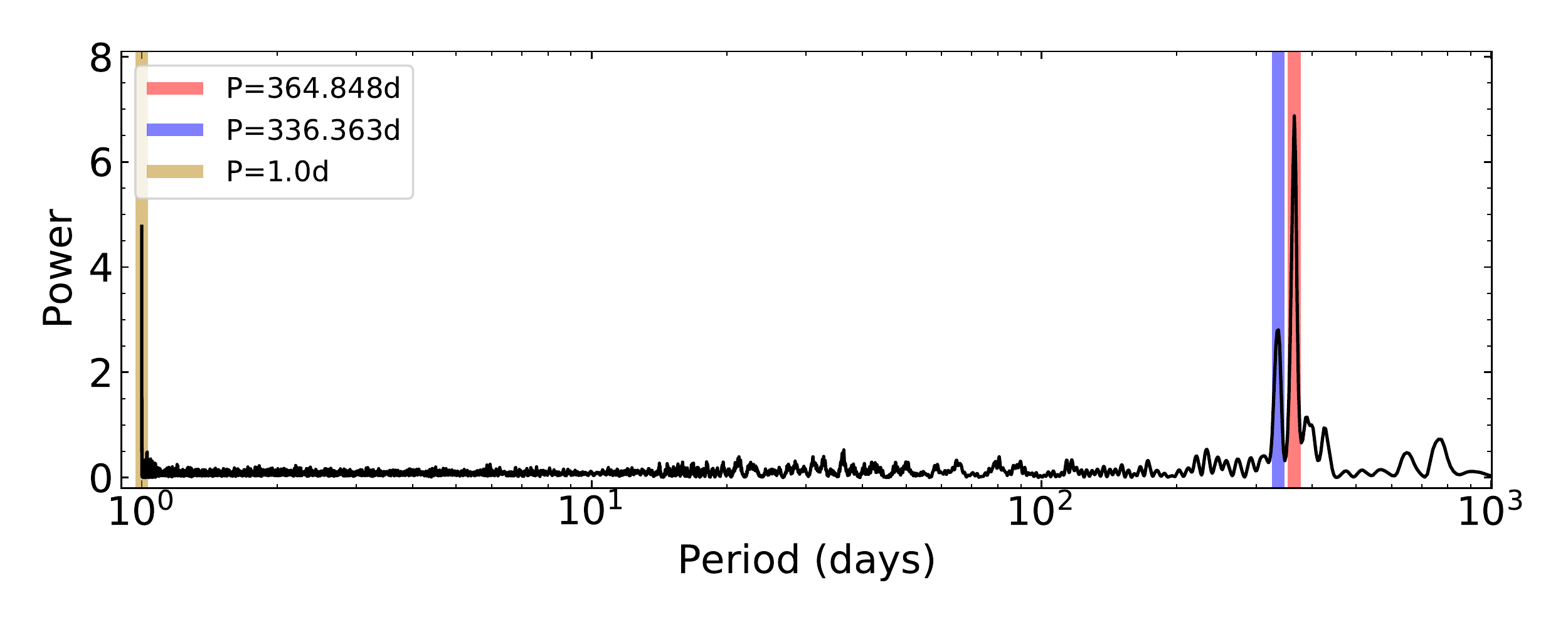}
    \caption{GLS periodogram of our observation sampling window function. The three prominent periods are shown using thick coloured lines with their respective values listed in the plot legend. The x-axis is shown in logarithmic scale.}
    \label{fig:Lombscargle_of_window_function_plot}
\end{figure}

 To confirm this, we compute the window function of our observations by setting our $y_{n}$ and $dy_{n}$ values equal to 1 for all times $t_{n}$ and run a classical periodogram on them with the values not pre-centered. The resulting periodogram, shown in figure  \ref{fig:Lombscargle_of_window_function_plot}, shows a peak at P=1d which is a common occurrence for ground-based surveys as the observations are done almost at the same time every other night. In addition to this, we detect a period of $364.8^{+5.2}_{-6.2}$d which perhaps is a result of the yearly gaps between the observing epochs and a shorter period of $336.4^{+5.7}_{-8.1}$d

\begin{table*}
    \renewcommand\thetable{3}
	\centering
	\caption{Observation log of the GJ 436 system observed by the NARVAL spectropolarimeter in 2016. The table contains the Signal-to-Noise ratio (SNR) and the H$\alpha$, NaI, CaII H\&K, CaI, HeI and CaII IRT activity indices along with their errors. The total exposure time for each observation spectra is  4$\times$700s. }
	\label{tab:Analysis_values_table}
	\resizebox{2\columnwidth}{!}{%
	\begin{tabular}{cccccccccccc} 
		\hline
		Date (2016) & HJD (2457000+) & UT (h:m:s) & S/N & H$\alpha$ & NaI & CaI & CaII H & HeI & IRT$_{1}$ & IRT$_{2}$ & IRT$_{3}$\\
		\hline
		\hline
		Mar 16 & 464.49670 & 23:49:19 & 203 & 0.4109 ± 0.0010 & 0.2144 ± 0.0022 & 0.3346 ± 0.0012 & 1.2210 ± 0.0960 & 0.4677 ± 0.0041 & 0.3021 ± 0.0007 & 0.2323 ± 0.0008 & 0.2156 ± 0.0006\\ 
        Mar 18 & 465.52978 & 00:36:58 & 204 & 0.4101 ± 0.0010 & 0.2300 ± 0.0022 & 0.3354 ± 0.0011 & 1.0388 ± 0.0927 & 0.4867 ± 0.0041 & 0.3083 ± 0.0007 & 0.2329 ± 0.0008 & 0.2158 ± 0.0006\\
        Mar 20 & 468.47996 & 23:25:14 & 274 & 0.4096 ± 0.0007 & 0.2159 ± 0.0013 & 0.3326 ± 0.0008 & 1.1705 ± 0.0499 & 0.4799 ± 0.0026 & 0.3032 ± 0.0005 & 0.2387 ± 0.0006 & 0.2171 ± 0.0004\\
        Apr 18 & 497.40198 & 21:33:54 & 235 & 0.4102 ± 0.0008 & 0.2148 ± 0.0016 & 0.3385 ± 0.0009 & 1.1727 ± 0.0573 & 0.4791 ± 0.0032 & 0.3036 ± 0.0006 & 0.2465 ± 0.0007 & 0.2167 ± 0.0004\\
        May 02 & 511.46867 & 23:10:51 & 275 & 0.4145 ± 0.0007 & 0.2110 ± 0.0013 & 0.3335 ± 0.0008 & 1.1942 ± 0.0444 & 0.4750 ± 0.0026 & 0.3068 ± 0.0005 & 0.2391 ± 0.0006 & 0.2163 ± 0.0004\\
        May 03 & 512.48383 & 23:32:45 & 227 & 0.4102 ± 0.0009 & 0.2176 ± 0.0018 & 0.3382 ± 0.0010 & 1.0720 ± 0.0650 & 0.4774 ± 0.0034 & 0.3061 ± 0.0006 & 0.2376 ± 0.0007 & 0.2165 ± 0.0005\\
        May 04 & 513.39868 & 21:30:12 & 234 & 0.4061 ± 0.0008 & 0.2141 ± 0.0017 & 0.3340 ± 0.0010 & 1.1243 ± 0.0631 & 0.4725 ± 0.0033 & 0.3068 ± 0.0006 & 0.2381 ± 0.0007 & 0.2169 ± 0.0004\\
        May 11 & 520.35975 & 20:34:43 & 204 & 0.4100 ± 0.0010 & 0.2051 ± 0.0020 & 0.3346 ± 0.0011 & 1.1530 ± 0.0763 & 0.4846 ± 0.0039 & 0.3026 ± 0.0007 & 0.2283 ± 0.0008 & 0.2224 ± 0.0005\\
        May 16 & 525.36718 & 20:45:50 & 273 & 0.4080 ± 0.0007 & 0.2122 ± 0.0013 & 0.3325 ± 0.0008 & 1.2993 ± 0.0458 & 0.4795 ± 0.0026 & 0.3031 ± 0.0005 & 0.2315 ± 0.0006 & 0.2216 ± 0.0004\\
        May 17 & 526.36784 & 20:46:53 & 216 & 0.4073 ± 0.0009 & 0.2212 ± 0.0018 & 0.3338 ± 0.0010 & 0.9972 ± 0.0617 & 0.4774 ± 0.0034 & 0.3031 ± 0.0007 & 0.2308 ± 0.0007 & 0.2230 ± 0.0005\\
        May 20 & 529.38615 & 21:13:31 & 219 & 0.4078 ± 0.0009 & 0.2093 ± 0.0019 & 0.3334 ± 0.0011 & 1.1308 ± 0.0752 & 0.4756 ± 0.0036 & 0.3072 ± 0.0006 & 0.2352 ± 0.0007 & 0.2206 ± 0.0005\\
        May 23 & 532.37271 & 20:54:26 & 233 & 0.4092 ± 0.0008 & 0.2069 ± 0.0017 & 0.3322 ± 0.0010 & 1.1895 ± 0.0653 & 0.4821 ± 0.0032 & 0.3057 ± 0.0006 & 0.2382 ± 0.0007 & 0.2156 ± 0.0005\\
        Jun 02 & 542.37484 & 20:58:28 & 219 & 0.4047 ± 0.0009 & 0.2192 ± 0.0019 & 0.3331 ± 0.0010 & 1.1213 ± 0.0690 & 0.4765 ± 0.0035 & 0.3051 ± 0.0006 & 0.2329 ± 0.0007 & 0.2206 ± 0.0005\\
        Jun 04 & 544.38322 & 21:10:44 & 212 & 0.4090 ± 0.0009 & 0.2100 ± 0.0019 & 0.3287 ± 0.0011 & 1.0549 ± 0.0731 & 0.4819 ± 0.0037 & 0.3075 ± 0.0007 & 0.2303 ± 0.0007 & 0.2245 ± 0.0005\\
        Jun 07 & 547.37509 & 20:59:19 & 272 & 0.4045 ± 0.0007 & 0.2060 ± 0.0013 & 0.3412 ± 0.0008 & 1.3019 ± 0.0517 & 0.4744 ± 0.0027 & 0.3078 ± 0.0005 & 0.2387 ± 0.0006 & 0.2227 ± 0.0004\\
        Jun 08 & 548.37866 & 21:04:34 & 289 & 0.4042 ± 0.0007 & 0.2058 ± 0.0012 & 0.3326 ± 0.0008 & 1.2250 ± 0.0410 & 0.4756 ± 0.0025 & 0.3064 ± 0.0005 & 0.2332 ± 0.0006 & 0.2218 ± 0.0003\\
		\hline
	\end{tabular}%
	}
\end{table*}

\begin{table*}
    \renewcommand\thetable{4}
	\centering
	\caption{Observation log of the GJ 436 system observed by HARPS from 2006 to 2020. The table contains the total exposure time $t_{exp}$ (s), median Signal-to-Noise ratio (SNR) and the H$\alpha$, NaI, CaII H\&K, CaI and HeI activity indices along with their errors}
	\label{tab:HARPS_analysis_table}
	\resizebox{2\columnwidth}{!}{%
	\begin{tabular}{cccccccccc}
		\hline
		Date & BJD (2450000+) & UT (h:m:s) & $t_{exp}$(s) & S/N & H$\alpha$ & NaI & CaII & CaI & HeI\\
		\hline
		\hline
		2006-01-25 & 3760.835 & 07:49:14.289 & 900.0017 & 22.25 & 0.4074±0.0008 & 0.1803±0.0016 & 0.5040±0.0166 & 0.3381±0.0011 & 0.4845±0.0028\\
        2006-01-26 & 3761.840 & 07:55:59.746 & 900.0012 & 21.85 & 0.4133±0.0008 & 0.1758±0.0015 & 0.6016±0.0180 & 0.3404±0.0011 & 0.4822±0.0028\\
        2006-01-27 & 3762.827 & 07:37:50.432 & 900.0006 & 31.55 & 0.4118±0.0006 & 0.1775±0.0011 & 0.5990±0.0102 & 0.3400±0.0008 & 0.4733±0.0020\\
        2006-01-28 & 3763.841 & 07:57:34.268 & 900.0005 & 31.70 & 0.4143±0.0006 & 0.1765±0.0011 & 0.5542±0.0099 & 0.3420±0.0008 & 0.4775±0.0020\\
        2006-01-30 & 3765.803 & 07:02:58.125 & 900.0012 & 28.60 & 0.4110±0.0007 & 0.1738±0.0012 & 0.5730±0.0117 & 0.3430±0.0008 & 0.4765±0.0022\\
        2006-02-19 & 3785.760 & 05:59:45.661 & 900.0025 & 29.40 & 0.4102±0.0006 & 0.1823±0.0012 & 0.5527±0.0107 & 0.3422±0.0008 & 0.4802±0.0022\\
        2006-02-22 & 3788.779 & 06:26:37.814 & 900.0009 & 32.15 & 0.4083±0.0006 & 0.1827±0.0011 & 0.5920±0.0092 & 0.3393±0.0008 & 0.4806±0.0020\\
        2007-01-22 & 4122.842 & 07:59:34.392 & 900.0001 & 30.70 & 0.4139±0.0006 & 0.1797±0.0011 & 0.5892±0.0094 & 0.3382±0.0008 & 0.4837±0.0020\\
        2007-02-04 & 4135.832 & 07:43:32.169 & 900.0017 & 31.05 & 0.4055±0.0006 & 0.1818±0.0011 & 0.5882±0.0091 & 0.3385±0.0008 & 0.4794±0.0021\\
        2007-02-09 & 4140.821 & 07:27:50.165 & 900.0034 & 24.65 & 0.4122±0.0008 & 0.1858±0.0014 & 0.5660±0.0123 & 0.3374±0.0010 & 0.4807±0.0025\\
        2007-02-11 & 4142.830 & 07:39:59.325 & 900.0021 & 30.40 & 0.4106±0.0006 & 0.1842±0.0011 & 0.6034±0.0095 & 0.3383±0.0008 & 0.4812±0.0020\\
        2007-03-07 & 4166.750 & 05:44:40.764 & 899.9954 & 20.50 & 0.4116±0.0009 & 0.1847±0.0017 & 0.5995±0.0168 & 0.3408±0.0011 & 0.4883±0.0029\\
        2007-03-13 & 4172.742 & 05:33:24.847 & 899.9972 & 25.75 & 0.4102±0.0007 & 0.1831±0.0013 & 0.5836±0.0120 & 0.3402±0.0009 & 0.4904±0.0024\\
        2007-04-04 & 4194.708 & 04:45:48.825 & 899.9961 & 18.00 & 0.4138±0.0010 & 0.1923±0.0019 & 0.6751±0.0194 & 0.3374±0.0012 & 0.4828±0.0032\\
        2007-04-07 & 4197.677 & 04:00:25.671 & 900.0021 & 21.60 & 0.4144±0.0008 & 0.1860±0.0015 & 0.6153±0.0153 & 0.3403±0.0010 & 0.4759±0.0027\\
        2007-04-09 & 4199.670 & 03:51:11.405 & 900.0019 & 26.95 & 0.4114±0.0007 & 0.1568±0.0011 & 0.5768±0.0113 & 0.3362±0.0009 & 0.4814±0.0023\\
        2007-04-12 & 4202.663 & 03:41:50.693 & 899.9963 & 27.90 & 0.4204±0.0007 & 0.1863±0.0012 & 0.6834±0.0108 & 0.3376±0.0008 & 0.4832±0.0022\\
        2007-05-08 & 4228.582 & 01:47:05.634 & 899.9962 & 23.95 & 0.4095±0.0007 & 0.1805±0.0014 & 0.6053±0.0124 & 0.3370±0.0010 & 0.4789±0.0025\\
        2007-05-09 & 4230.485 & 23:27:20.751 & 899.9945 & 19.60 & 0.4113±0.0009 & 0.1831±0.0016 & 0.6404±0.0171 & 0.3352±0.0011 & 0.4795±0.0030\\
        2007-05-09 & 4230.494 & 23:43:27.584 & 599.9964 & 17.10 & 0.4128±0.0010 & 0.1846±0.0019 & 0.6366±0.0196 & 0.3360±0.0013 & 0.4771±0.0033\\
        2007-05-09 & 4230.501 & 23:54:54.201 & 300.0005 & 10.70 & 0.4106±0.0015 & 0.1759±0.0028 & 0.8489±0.0374 & 0.3409±0.0019 & 0.4755±0.0049\\
        2007-05-10 & 4230.504 & 00:00:25.159 & 300.0005 & 11.10 & 0.4099±0.0014 & 0.1752±0.0027 & 0.5973±0.0341 & 0.3388±0.0018 & 0.4762±0.0048\\
        2007-05-10 & 4230.508 & 00:05:55.997 & 300.0026 & 12.40 & 0.4077±0.0013 & 0.1779±0.0025 & 0.5534±0.0295 & 0.3348±0.0017 & 0.4783±0.0043\\
        2007-05-10 & 4230.512 & 00:11:26.685 & 300.0052 & 13.10 & 0.4087±0.0012 & 0.1749±0.0023 & 0.5295±0.0278 & 0.3367±0.0016 & 0.4815±0.0041\\
        2007-05-10 & 4230.516 & 00:16:57.222 & 300.0005 & 11.60 & 0.4140±0.0014 & 0.1782±0.0026 & 0.5631±0.0328 & 0.3386±0.0018 & 0.4820±0.0046\\
        2007-05-10 & 4230.520 & 00:22:28.100 & 300.0007 & 11.80 & 0.4119±0.0014 & 0.1727±0.0025 & 0.6076±0.0312 & 0.3365±0.0018 & 0.4768±0.0046\\
        2007-05-10 & 4230.524 & 00:27:58.888 & 300.0001 &  6.95 & 0.4122±0.0021 & 0.1605±0.0040 & 0.6579±0.0577 & 0.3399±0.0028 & 0.4845±0.0070\\
        2007-05-10 & 4230.527 & 00:33:29.836 & 299.9996 &  9.60 & 0.4098±0.0016 & 0.1687±0.0030 & 0.5521±0.0393 & 0.3419±0.0021 & 0.4770±0.0054\\
        2007-05-10 & 4230.531 & 00:39:00.604 & 300.0049 & 11.80 & 0.4084±0.0014 & 0.1733±0.0026 & 0.7377±0.0319 & 0.3393±0.0018 & 0.4752±0.0046\\
        2007-05-10 & 4230.535 & 00:44:31.482 & 299.9988 & 13.55 & 0.4119±0.0012 & 0.1817±0.0024 & 0.6330±0.0256 & 0.3355±0.0016 & 0.4759±0.0041\\
        2007-05-10 & 4230.539 & 00:50:02.370 & 300.0036 & 13.90 & 0.4121±0.0012 & 0.1800±0.0023 & 0.6145±0.0249 & 0.3391±0.0016 & 0.4796±0.0040\\
        2007-05-10 & 4230.543 & 00:55:33.418 & 300.0006 & 14.05 & 0.4117±0.0012 & 0.1785±0.0023 & 0.6778±0.0251 & 0.3409±0.0016 & 0.4862±0.0040\\
        2007-05-10 & 4230.546 & 01:01:04.366 & 300.0007 & 13.90 & 0.4093±0.0012 & 0.1784±0.0023 & 0.6504±0.0260 & 0.3382±0.0016 & 0.4802±0.0041\\
        2007-05-10 & 4230.550 & 01:06:35.533 & 300.0032 & 14.00 & 0.4097±0.0012 & 0.1786±0.0022 & 0.6397±0.0253 & 0.3411±0.0016 & 0.4785±0.0040\\
        2007-05-10 & 4230.554 & 01:12:06.511 & 299.9994 & 13.75 & 0.4135±0.0012 & 0.1802±0.0023 & 0.6856±0.0269 & 0.3373±0.0016 & 0.4918±0.0042\\
        2007-05-10 & 4230.558 & 01:17:37.779 & 299.9975 & 15.25 & 0.4121±0.0011 & 0.1774±0.0021 & 0.6274±0.0228 & 0.3390±0.0014 & 0.4756±0.0037\\
        2007-05-10 & 4230.562 & 01:23:08.497 & 299.9991 & 15.80 & 0.4095±0.0011 & 0.1764±0.0020 & 0.7011±0.0224 & 0.3394±0.0014 & 0.4772±0.0036\\
        2007-05-10 & 4230.566 & 01:28:39.555 & 300.0008 & 14.70 & 0.4103±0.0012 & 0.1830±0.0022 & 0.6819±0.0238 & 0.3382±0.0015 & 0.4789±0.0038\\
        2007-05-10 & 4230.569 & 01:34:10.553 & 300.0056 & 14.65 & 0.4163±0.0012 & 0.1847±0.0022 & 0.6955±0.0238 & 0.3405±0.0015 & 0.4831±0.0039\\
        2007-05-10 & 4230.573 & 01:39:41.541 & 300.0054 & 17.45 & 0.4174±0.0010 & 0.1858±0.0019 & 0.7150±0.0191 & 0.3389±0.0013 & 0.4776±0.0033\\
        2007-05-10 & 4230.577 & 01:45:12.469 & 300.0042 & 17.10 & 0.4159±0.0010 & 0.1827±0.0019 & 0.7362±0.0204 & 0.3386±0.0013 & 0.4704±0.0033\\
        2007-05-10 & 4230.581 & 01:50:43.807 & 300.0002 & 14.35 & 0.4100±0.0012 & 0.1815±0.0022 & 0.7038±0.0245 & 0.3372±0.0015 & 0.4817±0.0039\\
        2007-05-10 & 4230.585 & 01:56:14.615 & 299.9981 & 16.95 & 0.4092±0.0010 & 0.1880±0.0020 & 0.6465±0.0196 & 0.3384±0.0013 & 0.4827±0.0035\\
        2007-05-10 & 4230.589 & 02:01:45.983 & 299.9997 & 17.05 & 0.4099±0.0010 & 0.1788±0.0019 & 0.6350±0.0193 & 0.3397±0.0013 & 0.4759±0.0034\\
        2007-05-10 & 4230.592 & 02:07:17.221 & 299.9982 & 14.80 & 0.4139±0.0012 & 0.1768±0.0021 & 0.5695±0.0230 & 0.3388±0.0015 & 0.4798±0.0038\\
        2007-05-10 & 4230.596 & 02:12:48.108 & 299.9979 & 15.70 & 0.4099±0.0011 & 0.1787±0.0020 & 0.5888±0.0209 & 0.3380±0.0014 & 0.4792±0.0037\\
        2007-05-10 & 4230.600 & 02:18:18.966 & 300.0016 & 12.70 & 0.409±0.00130 & 0.1722±0.0024 & 0.6108±0.0280 & 0.3378±0.0017 & 0.4764±0.0044\\
        2007-05-10 & 4230.604 & 02:23:49.754 & 300.0013 & 13.25 & 0.4077±0.0013 & 0.1810±0.0024 & 0.5406±0.0258 & 0.3372±0.0016 & 0.4809±0.0042\\
        2007-05-10 & 4230.608 & 02:29:20.712 & 299.9999 & 11.30 & 0.4066±0.0014 & 0.1747±0.0027 & 0.6489±0.0331 & 0.3374±0.0019 & 0.4798±0.0048\\
        2007-05-10 & 4230.612 & 02:34:51.480 & 299.9953 & 11.05 & 0.4113±0.0014 & 0.1775±0.0028 & 0.5282±0.0343 & 0.3366±0.0019 & 0.4819±0.0049\\
		\hline
	\end{tabular}%
	}
\end{table*}

\begin{table*}
	\centering
	\resizebox{2\columnwidth}{!}{%
	\begin{tabular}{cccccccccc}
		\hline
		2007-05-10 & 4230.615 & 02:40:22.158 & 300.0006 &  9.60 & 0.4137±0.0016 & 0.1723±0.0031 & 0.6209±0.0405 & 0.3361±0.0021 & 0.4799±0.0055\\
        2007-05-10 & 4230.619 & 02:45:52.966 & 299.9987 &  9.95 & 0.4147±0.0016 & 0.1700±0.0030 & 0.6503±0.0409 & 0.3402±0.0021 & 0.4810±0.0053\\
        2007-05-10 & 4230.623 & 02:51:23.774 & 299.9947 & 10.25 & 0.412±0.00160 & 0.1721±0.0029 & 0.6887±0.0381 & 0.3397±0.0020 & 0.4854±0.0052\\
        2007-05-10 & 4230.627 & 02:56:54.752 & 299.9947 & 13.45 & 0.4125±0.0012 & 0.1826±0.0024 & 0.6512±0.0274 & 0.3382±0.0016 & 0.4727±0.0041\\
        2007-05-10 & 4230.631 & 03:02:25.609 & 299.9954 & 14.00 & 0.4083±0.0012 & 0.1783±0.0022 & 0.6801±0.0260 & 0.3368±0.0015 & 0.4833±0.0040\\
        2007-05-10 & 4230.635 & 03:07:56.437 & 299.9947 & 15.95 & 0.4125±0.0011 & 0.1807±0.0020 & 0.6487±0.0226 & 0.3371±0.0014 & 0.4809±0.0036\\
        2007-05-10 & 4230.639 & 03:13:53.497 & 299.9947 & 15.75 & 0.4119±0.0011 & 0.1799±0.0020 & 0.6100±0.0226 & 0.3370±0.0014 & 0.4799±0.0036\\
        2007-05-10 & 4230.643 & 03:19:24.165 & 299.9956 & 16.40 & 0.4097±0.0011 & 0.1780±0.0019 & 0.6151±0.0212 & 0.3376±0.0013 & 0.4763±0.0034\\
        2007-05-10 & 4230.646 & 03:24:55.132 & 299.9995 & 15.95 & 0.4089±0.0011 & 0.1800±0.0020 & 0.5529±0.0214 & 0.3379±0.0014 & 0.4825±0.0036\\
        2007-05-10 & 4230.650 & 03:30:26.120 & 299.9947 & 16.55 & 0.4106±0.0010 & 0.1754±0.0019 & 0.5685±0.0210 & 0.3356±0.0013 & 0.4761±0.0034\\
        2007-05-10 & 4230.654 & 03:35:57.118 & 299.9945 & 15.60 & 0.4075±0.0011 & 0.1808±0.0020 & 0.5998±0.0235 & 0.3370±0.0014 & 0.4765±0.0035\\
        2007-05-10 & 4230.658 & 03:41:28.076 & 299.9956 & 14.60 & 0.4119±0.0011 & 0.1832±0.0021 & 0.7136±0.0269 & 0.3395±0.0014 & 0.4776±0.0036\\
        2007-05-14 & 4234.553 & 01:05:12.938 & 899.9974 & 19.55 & 0.4142±0.0009 & 0.1843±0.0017 & 0.6804±0.0176 & 0.3406±0.0011 & 0.4825±0.0030\\
        2007-06-02 & 4253.545 & 00:49:00.762 & 1799.9950 & 22.75 & 0.4089±0.0008 & 0.1838±0.0015 & 0.5051±0.0139 & 0.3389±0.0010 & 0.4793±0.0026\\
		2007-06-03 & 4254.511 & 00:01:11.437 & 1799.9976 & 19.45 & 0.4073±0.0009 & 0.1617±0.0015 & 0.5947±0.0181 & 0.3380±0.0012 & 0.4766±0.0030\\
        2007-06-04 & 4255.516 & 00:01:55.802 & 1999.9987 & 20.15 & 0.4074±0.0009 & 0.1806±0.0016 & 0.5530±0.0169 & 0.3387±0.0011 & 0.4733±0.0029\\
        2007-06-07 & 4259.487 & 23:25:53.536 & 1800.0005 & 33.95 & 0.4077±0.0005 & 0.1766±0.0010 & 0.5268±0.0077 & 0.3392±0.0007 & 0.4814±0.0018\\
        2007-07-09 & 4291.488 & 23:39:41.130 & 899.9997 & 10.85 & 0.4150±0.0014 & 0.1745±0.0027 & 0.6260±0.0534 & 0.3380±0.0018 & 0.4762±0.0048\\
        2007-07-10 & 4292.474 & 23:19:57.798 & 899.9983 & 27.65 & 0.4099±0.0007 & 0.1818±0.0012 & 0.6225±0.0109 & 0.3362±0.0009 & 0.4755±0.0022\\
        2007-07-11 & 4293.451 & 22:46:11.010 & 899.9988 & 34.25 & 0.4126±0.0005 & 0.1841±0.0010 & 0.6210±0.0082 & 0.3387±0.0007 & 0.4791±0.0017\\
        2007-07-12 & 4294.449 & 22:44:03.574 & 900.0002 & 23.85 & 0.4089±0.0007 & 0.1852±0.0014 & 0.5741±0.0123 & 0.3381±0.0009 & 0.4713±0.0024\\
        2007-07-14 & 4296.476 & 23:23:09.149 & 900.0002 & 16.35 & 0.4062±0.0010 & 0.1838±0.0019 & 0.6394±0.0226 & 0.3399±0.0013 & 0.4777±0.0034\\
        2007-07-15 & 4297.453 & 22:49:24.843 & 899.9997 & 26.00 & 0.4082±0.0007 & 0.1817±0.0012 & 0.5832±0.0111 & 0.3393±0.0009 & 0.4794±0.0023\\
		2008-01-13 & 4478.852 & 08:14:31.312 & 900.0000 & 27.70 & 0.4170±0.0007 & 0.1755±0.0012 & 0.5035±0.0097 & 0.3357±0.0008 & 0.4802±0.0022\\
        2008-01-13 & 4478.878 & 08:51:43.283 & 900.0000 & 28.50 & 0.4175±0.0007 & 0.1790±0.0012 & 0.5493±0.0095 & 0.3364±0.0008 & 0.4786±0.0021\\
        2008-01-14 & 4479.865 & 08:32:55.089 & 900.0000 & 23.35 & 0.4171±0.0008 & 0.1777±0.0014 & 0.5532±0.0127 & 0.3348±0.0010 & 0.4808±0.0026\\
        2008-01-14 & 4479.875 & 08:48:25.929 & 900.0000 & 24.75 & 0.4168±0.0008 & 0.1792±0.0013 & 0.5319±0.0115 & 0.3353±0.0009 & 0.4864±0.0025\\
        2008-01-15 & 4480.858 & 08:24:14.411 & 900.0000 & 26.85 & 0.4166±0.0007 & 0.1748±0.0012 & 0.5144±0.0103 & 0.3347±0.0009 & 0.4832±0.0023\\
        2008-01-15 & 4480.870 & 08:39:45.762 & 900.0000 & 22.20 & 0.4157±0.0008 & 0.1719±0.0014 & 0.5377±0.0134 & 0.3336±0.0010 & 0.4783±0.0027\\
        2008-01-16 & 4481.862 & 08:29:09.201 & 900.0000 & 26.80 & 0.4190±0.0007 & 0.1663±0.0012 & 0.5672±0.0106 & 0.3359±0.0009 & 0.4746±0.0023\\
        2008-01-16 & 4481.873 & 08:44:40.662 & 900.0000 & 28.55 & 0.4188±0.0007 & 0.1732±0.0011 & 0.5863±0.0098 & 0.3351±0.0008 & 0.4736±0.0021\\
        2008-01-17 & 4482.853 & 08:15:59.848 & 900.0000 & 23.45 & 0.4161±0.0008 & 0.1637±0.0013 & 0.5629±0.0123 & 0.3364±0.0010 & 0.4768±0.0026\\
        2008-01-17 & 4482.864 & 08:31:30.948 & 900.0000 & 23.80 & 0.4167±0.0008 & 0.1712±0.0013 & 0.5668±0.0122 & 0.3340±0.0010 & 0.4718±0.0025\\
        2008-01-18 & 4483.852 & 08:13:36.738 & 900.0000 & 25.30 & 0.4202±0.0007 & 0.1742±0.0013 & 0.5877±0.0117 & 0.3364±0.0009 & 0.4761±0.0024\\
        2008-01-18 & 4483.863 & 08:29:08.038 & 900.0000 & 28.15 & 0.4217±0.0007 & 0.1743±0.0011 & 0.5809±0.0101 & 0.3355±0.0008 & 0.4775±0.0021\\
        2008-01-20 & 4485.857 & 08:20:53.267 & 900.0000 & 25.40 & 0.4193±0.0007 & 0.1795±0.0013 & 0.5849±0.0118 & 0.3362±0.0009 & 0.4820±0.0024\\
        2008-01-20 & 4485.868 & 08:36:24.928 & 900.0000 & 25.50 & 0.4160±0.0007 & 0.1775±0.0013 & 0.5293±0.0116 & 0.3378±0.0009 & 0.4788±0.0024\\
        2008-01-21 & 4486.854 & 08:16:30.074 & 900.0000 & 23.40 & 0.4156±0.0008 & 0.1786±0.0014 & 0.5554±0.0131 & 0.3366±0.0010 & 0.4776±0.0026\\
        2008-01-21 & 4486.864 & 08:32:01.905 & 900.0000 & 21.10 & 0.4144±0.0009 & 0.1798±0.0016 & 0.5586±0.0153 & 0.3368±0.0011 & 0.4793±0.0028\\
        2008-01-22 & 4487.851 & 08:12:19.811 & 900.0000 & 18.95 & 0.4125±0.0009 & 0.1749±0.0017 & 0.4912±0.0167 & 0.3377±0.0012 & 0.4790±0.0031\\
        2008-01-22 & 4487.862 & 08:27:53.061 & 900.0000 & 23.30 & 0.4146±0.0008 & 0.1756±0.0014 & 0.5229±0.0128 & 0.3409±0.0010 & 0.4802±0.0026\\
        2008-01-23 & 4488.854 & 08:17:27.757 & 900.0000 & 21.65 & 0.4176±0.0008 & 0.1818±0.0015 & 0.5368±0.0141 & 0.3390±0.0011 & 0.4762±0.0027\\
        2008-02-26 & 4522.838 & 07:51:57.296 & 900.0000 & 17.80 & 0.4101±0.0009 & 0.1833±0.0018 & 0.6189±0.0192 & 0.3414±0.0012 & 0.4767±0.0031\\
        2008-02-27 & 4523.789 & 06:40:48.149 & 900.0000 & 22.90 & 0.4196±0.0008 & 0.1823±0.0014 & 0.6434±0.0129 & 0.3405±0.0010 & 0.4911±0.0026\\
        2008-02-28 & 4524.789 & 06:41:55.545 & 900.0000 & 31.30 & 0.4085±0.0006 & 0.1782±0.0011 & 0.5228±0.0082 & 0.3381±0.0008 & 0.4791±0.0020\\
        2008-02-29 & 4525.812 & 07:15:20.355 & 900.0000 & 21.20 & 0.4098±0.0008 & 0.1745±0.0015 & 0.5269±0.0143 & 0.3398±0.0011 & 0.4793±0.0027\\
        2008-03-01 & 4526.763 & 06:03:17.367 & 900.0000 & 30.85 & 0.4119±0.0006 & 0.1761±0.0011 & 0.5449±0.0085 & 0.3391±0.0008 & 0.4806±0.0020\\
        2008-03-02 & 4527.751 & 05:46:18.300 & 900.0000 & 26.60 & 0.4135±0.0007 & 0.1749±0.0012 & 0.5139±0.0103 & 0.3384±0.0009 & 0.4825±0.0022\\
        2008-03-03 & 4528.717 & 04:56:50.095 & 900.0000 & 28.75 & 0.4171±0.0006 & 0.1669±0.0010 & 0.6229±0.0096 & 0.3402±0.0008 & 0.4821±0.0021\\
        2008-03-04 & 4529.812 & 07:13:47.000 & 900.0000 & 25.40 & 0.4157±0.0007 & 0.1759±0.0012 & 0.5258±0.0113 & 0.3411±0.0009 & 0.4846±0.0023\\
        2008-03-05 & 4530.777 & 06:23:27.951 & 900.0000 & 26.35 & 0.4131±0.0007 & 0.1765±0.0012 & 0.5471±0.0107 & 0.3422±0.0009 & 0.4823±0.0022\\
        2008-03-23 & 4548.703 & 04:29:09.320 & 1800.0000 & 31.95 & 0.4120±0.0006 & 0.1832±0.0010 & 0.6551±0.0086 & 0.3374±0.0007 & 0.4805±0.0019\\
        2008-03-24 & 4549.697 & 04:23:08.332 & 1800.0000 & 20.80 & 0.4070±0.0009 & 0.1870±0.0016 & 0.5574±0.0142 & 0.3351±0.0011 & 0.4801±0.0029\\
        2008-03-26 & 4551.718 & 04:50:59.515 & 1800.0000 & 27.15 & 0.4134±0.0007 & 0.1816±0.0012 & 0.6116±0.0106 & 0.3387±0.0008 & 0.4783±0.0022\\
        2008-03-28 & 4553.697 & 04:20:30.367 & 1800.0000 & 39.50 & 0.4098±0.0005 & 0.1779±0.0008 & 0.5803±0.0064 & 0.3392±0.0006 & 0.4839±0.0015\\
        2008-04-01 & 4557.658 & 03:31:59.932 & 900.0000 & 21.30 & 0.4120±0.0008 & 0.1820±0.0015 & 0.5680±0.0154 & 0.3381±0.0011 & 0.4825±0.0027\\
        2008-04-06 & 4562.644 & 03:13:07.077 & 900.0000 & 25.55 & 0.4191±0.0007 & 0.1848±0.0013 & 0.6829±0.0111 & 0.3393±0.0009 & 0.4775±0.0023\\
        2008-04-08 & 4564.647 & 03:17:17.360 & 900.0000 & 18.40 & 0.4067±0.0009 & 0.1794±0.0017 & 0.5865±0.0166 & 0.3390±0.0012 & 0.4770±0.0031\\
        2008-04-11 & 4567.565 & 01:18:00.524 & 1200.0000 & 36.80 & 0.4135±0.0005 & 0.1801±0.0009 & 0.5887±0.0068 & 0.3370±0.0006 & 0.4823±0.0016\\
        2008-04-11 & 4567.734 & 05:21:29.227 & 1200.0000 & 31.85 & 0.4106±0.0005 & 0.1748±0.0010 & 0.5584±0.0088 & 0.3377±0.0007 & 0.4822±0.0018\\
        2008-04-12 & 4568.570 & 01:24:47.042 & 1200.0000 & 31.00 & 0.4121±0.0006 & 0.1811±0.0010 & 0.5568±0.0084 & 0.3381±0.0007 & 0.4826±0.0019\\
        2008-04-12 & 4568.691 & 04:18:39.251 & 1200.0000 & 27.65 & 0.4113±0.0007 & 0.1771±0.0011 & 0.5483±0.0099 & 0.3377±0.0008 & 0.4839±0.0021\\
        2008-04-13 & 4569.628 & 02:48:27.256 & 1200.0000 & 33.00 & 0.4115±0.0006 & 0.1764±0.0010 & 0.5453±0.0076 & 0.3371±0.0007 & 0.4818±0.0018\\
        2008-04-14 & 4570.637 & 03:01:26.706 & 1200.0000 & 29.00 & 0.4100±0.0006 & 0.1744±0.0011 & 0.5710±0.0092 & 0.3368±0.0008 & 0.4843±0.0021\\
        2008-04-15 & 4571.609 & 02:20:57.077 & 1200.0000 & 24.30 & 0.4100±0.0007 & 0.1817±0.0013 & 0.6037±0.0116 & 0.3367±0.0009 & 0.4827±0.0024\\
        2008-05-07 & 4593.551 & 01:02:54.710 & 900.0000 & 27.20 & 0.4097±0.0007 & 0.1800±0.0012 & 0.5546±0.0107 & 0.3366±0.0009 & 0.4840±0.0022\\
        2008-05-23 & 4610.478 & 23:20:22.079 & 900.0000 & 21.15 & 0.4089±0.0008 & 0.1771±0.0015 & 0.5964±0.0162 & 0.3379±0.0011 & 0.4791±0.0028\\
        2008-05-24 & 4610.562 & 01:21:06.631 & 900.0000 & 22.25 & 0.4077±0.0008 & 0.1779±0.0014 & 0.5571±0.0149 & 0.3387±0.0010 & 0.4753±0.0026\\
        2008-05-25 & 4611.557 & 01:13:01.347 & 900.0000 & 24.10 & 0.4074±0.0007 & 0.1773±0.0013 & 0.5630±0.0130 & 0.3370±0.0010 & 0.4834±0.0025\\
		\hline
	\end{tabular}%
	}
\end{table*}

\begin{table*}
	\centering
	\resizebox{2\columnwidth}{!}{%
	\begin{tabular}{cccccccccc}
		\hline
		2008-05-29 & 4616.476 & 23:17:49.454 & 900.0000 & 15.40 & 0.4099±0.0011 & 0.1759±0.0020 & 0.5896±0.0243 & 0.3339±0.0014 & 0.4764±0.0037\\
        2009-01-01 & 4832.876 & 08:50:43.355 & 900.0000 & 19.75 & 0.4099±0.0009 & 0.1875±0.0017 & 0.4952±0.0138 & 0.3357±0.0012 & 0.4773±0.0030\\
        2009-01-17 & 4848.871 & 08:40:56.016 & 900.0000 & 24.90 & 0.4116±0.0008 & 0.1791±0.0013 & 0.5695±0.0113 & 0.3368±0.0010 & 0.4752±0.0025\\
        2009-01-23 & 4854.846 & 08:05:36.145 & 900.0000 & 23.60 & 0.4147±0.0008 & 0.1821±0.0014 & 0.6078±0.0126 & 0.3386±0.0010 & 0.4806±0.0025\\
        2009-02-16 & 4878.789 & 06:41:42.107 & 900.0000 & 28.35 & 0.4130±0.0007 & 0.1764±0.0011 & 0.4927±0.0098 & 0.3409±0.0008 & 0.4783±0.0021\\
        2009-02-17 & 4879.791 & 06:44:41.282 & 900.0000 & 15.40 & 0.4117±0.0011 & 0.1775±0.0020 & 0.5336±0.0260 & 0.3406±0.0014 & 0.4819±0.0036\\
        2009-02-18 & 4880.797 & 06:53:32.974 & 900.0000 & 11.95 & 0.4151±0.0014 & 0.1770±0.0026 & 0.7077±0.0368 & 0.3396±0.0018 & 0.4942±0.0046\\
        2009-02-18 & 4880.812 & 07:14:56.353 & 900.0000 & 18.45 & 0.4172±0.0010 & 0.1768±0.0017 & 0.5535±0.0186 & 0.3410±0.0013 & 0.4861±0.0032\\
        2009-02-19 & 4881.787 & 06:38:37.801 & 900.0000 & 25.60 & 0.4140±0.0007 & 0.1766±0.0013 & 0.5375±0.0117 & 0.3413±0.0009 & 0.4821±0.0023\\
        2009-02-20 & 4882.784 & 06:33:25.497 & 900.0000 & 21.45 & 0.4123±0.0008 & 0.1767±0.0015 & 0.4987±0.0149 & 0.3405±0.0011 & 0.4764±0.0027\\
		2009-02-21 & 4883.794 & 06:48:10.721 & 900.0000 & 28.80 & 0.4126±0.0006 & 0.1784±0.0011 & 0.5113±0.0095 & 0.3395±0.0008 & 0.4819±0.0021\\
        2009-02-22 & 4884.773 & 06:18:11.456 & 900.0000 & 15.90 & 0.4107±0.0011 & 0.1769±0.0020 & 0.5065±0.0201 & 0.3376±0.0015 & 0.4825±0.0037\\
        2009-02-23 & 4885.792 & 06:45:47.206 & 900.0000 & 29.85 & 0.4126±0.0006 & 0.1792±0.0011 & 0.5290±0.0086 & 0.3379±0.0008 & 0.4788±0.0020\\
        2009-02-24 & 4886.769 & 06:12:47.455 & 900.0000 & 21.65 & 0.4114±0.0008 & 0.1772±0.0014 & 0.5510±0.0132 & 0.3405±0.0010 & 0.4902±0.0026\\
        2009-03-23 & 4913.688 & 04:15:54.065 & 900.0000 & 27.95 & 0.4063±0.0006 & 0.1812±0.0012 & 0.5356±0.0094 & 0.3372±0.0008 & 0.4717±0.0021\\
        2009-03-24 & 4914.708 & 04:44:20.901 & 900.0000 & 24.75 & 0.4059±0.0007 & 0.1812±0.0013 & 0.5933±0.0117 & 0.3377±0.0009 & 0.4783±0.0023\\
        2009-03-25 & 4915.687 & 04:14:59.071 & 900.0000 & 22.95 & 0.4065±0.0008 & 0.1801±0.0014 & 0.5389±0.0125 & 0.3384±0.0010 & 0.4740±0.0025\\
        2009-03-26 & 4916.692 & 04:22:10.575 & 900.0000 & 27.05 & 0.4052±0.0007 & 0.1764±0.0012 & 0.5701±0.0101 & 0.3387±0.0008 & 0.4727±0.0022\\
        2009-03-27 & 4917.680 & 04:04:13.404 & 900.0000 & 24.20 & 0.4087±0.0007 & 0.1802±0.0013 & 0.5709±0.0116 & 0.3391±0.0009 & 0.4757±0.0024\\
        2009-03-28 & 4918.727 & 05:12:44.335 & 900.0000 & 25.25 & 0.4091±0.0007 & 0.1781±0.0013 & 0.5841±0.0111 & 0.3369±0.0009 & 0.4760±0.0023\\
        2009-03-29 & 4919.662 & 03:39:11.570 & 900.0000 & 23.75 & 0.4121±0.0007 & 0.1820±0.0013 & 0.5875±0.0118 & 0.3390±0.0009 & 0.4800±0.0024\\
        2009-03-30 & 4920.707 & 04:43:16.572 & 900.0000 & 26.70 & 0.4111±0.0007 & 0.1821±0.0012 & 0.5583±0.0101 & 0.3380±0.0009 & 0.4793±0.0022\\
        2009-04-11 & 4932.643 & 03:11:18.807 & 900.0000 & 16.65 & 0.4095±0.0010 & 0.1811±0.0019 & 0.5416±0.0175 & 0.3371±0.0013 & 0.4872±0.0034\\
        2009-04-12 & 4933.638 & 03:04:47.794 & 900.0000 & 22.30 & 0.4077±0.0008 & 0.1775±0.0014 & 0.5602±0.0122 & 0.3392±0.0010 & 0.4748±0.0026\\
        2009-04-13 & 4934.626 & 02:47:59.458 & 900.0000 & 24.25 & 0.4079±0.0007 & 0.1688±0.0012 & 0.5700±0.0112 & 0.3384±0.0009 & 0.4792±0.0024\\
        2009-04-15 & 4936.632 & 02:56:56.447 & 900.0000 & 14.20 & 0.4051±0.0012 & 0.1763±0.0022 & 0.5988±0.0193 & 0.3413±0.0016 & 0.4771±0.0039\\
        2009-04-16 & 4937.611 & 02:27:03.710 & 900.0000 & 18.95 & 0.4081±0.0009 & 0.1816±0.0017 & 0.5524±0.0153 & 0.3375±0.0012 & 0.4771±0.0030\\
        2009-04-17 & 4938.623 & 02:43:32.285 & 900.0000 & 27.10 & 0.4064±0.0007 & 0.1797±0.0012 & 0.5556±0.0095 & 0.3384±0.0008 & 0.4768±0.0022\\
        2009-04-18 & 4939.639 & 03:06:34.539 & 900.0000 & 14.70 & 0.4085±0.0011 & 0.1785±0.0021 & 0.6309±0.0216 & 0.3366±0.0015 & 0.4789±0.0037\\
        2009-04-19 & 4940.614 & 02:31:35.787 & 900.0000 & 15.45 & 0.4071±0.0011 & 0.1822±0.0020 & 0.6253±0.0203 & 0.3364±0.0014 & 0.4722±0.0035\\
        2009-04-20 & 4941.622 & 02:43:00.570 & 900.0000 & 25.80 & 0.4031±0.0007 & 0.1781±0.0012 & 0.5901±0.0102 & 0.3360±0.0009 & 0.4796±0.0023\\
        2009-04-25 & 4946.604 & 02:17:43.389 & 900.0000 & 30.90 & 0.4074±0.0006 & 0.1799±0.0011 & 0.5507±0.0079 & 0.3377±0.0008 & 0.4793±0.0019\\
        2009-04-28 & 4949.582 & 01:46:07.435 & 900.0000 & 26.10 & 0.4056±0.0007 & 0.1821±0.0013 & 0.5672±0.0100 & 0.3378±0.0009 & 0.4714±0.0023\\
        2009-04-29 & 4950.583 & 01:47:55.171 & 900.0000 & 27.85 & 0.4082±0.0007 & 0.1830±0.0012 & 0.5761±0.0093 & 0.3358±0.0008 & 0.4769±0.0021\\
		2009-05-02 & 4953.588 & 01:54:52.741 & 900.0000 & 23.20 & 0.4035±0.0008 & 0.1594±0.0012 & 0.5286±0.0128 & 0.3375±0.0010 & 0.4780±0.0025\\
        2009-05-03 & 4954.583 & 01:47:54.873 & 900.0000 & 26.75 & 0.4042±0.0007 & 0.1640±0.0011 & 0.5361±0.0104 & 0.3399±0.0009 & 0.4750±0.0022\\
        2009-05-04 & 4955.582 & 01:45:25.155 & 900.0000 & 26.00 & 0.3985±0.0007 & 0.1766±0.0012 & 0.5283±0.0107 & 0.3384±0.0009 & 0.4761±0.0023\\
        2009-05-05 & 4956.583 & 01:48:12.726 & 900.0000 & 18.65 & 0.3962±0.0009 & 0.1787±0.0017 & 0.5301±0.0176 & 0.3385±0.0012 & 0.4691±0.0030\\
        2009-06-07 & 4990.498 & 23:49:42.435 & 899.9994 & 20.80 & 0.4063±0.0008 & 0.1812±0.0015 & 0.4462±0.0145 & 0.3373±0.0011 & 0.4814±0.0027\\
        2009-06-09 & 4991.510 & 00:07:46.413 & 900.0005 & 24.70 & 0.4061±0.0008 & 0.1817±0.0013 & 0.4472±0.0111 & 0.3375±0.0010 & 0.4828±0.0024\\
        2009-06-10 & 4993.465 & 23:02:22.549 & 899.9991 & 26.40 & 0.4060±0.0007 & 0.1807±0.0012 & 0.5055±0.0104 & 0.3369±0.0009 & 0.4779±0.0022\\
        2010-03-17 & 5272.702 & 04:35:49.928 & 899.9998 & 21.05 & 0.4093±0.0008 & 0.1778±0.0015 & 0.4912±0.0149 & 0.3386±0.0011 & 0.4769±0.0027\\
        2010-03-20 & 5275.707 & 04:43:15.807 & 900.0068 & 16.30 & 0.4089±0.0010 & 0.1820±0.0019 & 0.5149±0.0227 & 0.3399±0.0013 & 0.4734±0.0033\\
        2010-03-22 & 5277.711 & 04:48:38.792 & 899.9990 & 16.30 & 0.4173±0.0010 & 0.1808±0.0018 & 0.7036±0.0223 & 0.3392±0.0012 & 0.4816±0.0032\\
        2010-03-23 & 5278.703 & 04:36:33.665 & 899.9985 & 11.65 & 0.4085±0.0014 & 0.1752±0.0025 & 0.4999±0.0365 & 0.3368±0.0018 & 0.4836±0.0045\\
        2010-03-24 & 5279.684 & 04:09:33.598 & 899.9987 & 19.40 & 0.4083±0.0009 & 0.1815±0.0016 & 0.4946±0.0171 & 0.3376±0.0012 & 0.4757±0.0029\\
        2010-03-25 & 5280.700 & 04:33:03.332 & 900.0001 & 19.20 & 0.4097±0.0009 & 0.1807±0.0017 & 0.4938±0.0170 & 0.3370±0.0012 & 0.4760±0.0030\\
        2010-03-28 & 5283.677 & 04:00:55.642 & 900.0064 & 26.35 & 0.4120±0.0007 & 0.1792±0.0012 & 0.5204±0.0095 & 0.3387±0.0009 & 0.4828±0.0022\\
        2010-04-01 & 5287.655 & 03:29:21.475 & 899.9999 & 21.65 & 0.4096±0.0008 & 0.1856±0.0015 & 0.5895±0.0131 & 0.3363±0.0010 & 0.4778±0.0026\\
        2010-04-06 & 5292.649 & 03:22:20.916 & 900.0069 &  8.00 & 0.4046±0.0018 & 0.1870±0.0037 & 0.4311±0.0376 & 0.3381±0.0024 & 0.4704±0.0059\\
        2020-01-28 & 8876.843 & 08:00:12.543 & 899.9976 & 25.05 & 0.4169±0.0007 & 0.1909±0.0012 & 0.5824±0.0096 & 0.3384±0.0009 & 0.4783±0.0022\\
        2020-01-31 & 8879.843 & 07:59:54.501 & 899.9977 & 17.35 & 0.4131±0.0009 & 0.1914±0.0017 & 0.6253±0.0140 & 0.3387±0.0012 & 0.4778±0.0029\\
        2020-02-01 & 8880.832 & 07:44:08.123 & 899.9977 & 20.30 & 0.4083±0.0008 & 0.1855±0.0014 & 0.6500±0.0123 & 0.3392±0.0010 & 0.4852±0.0026\\
        2020-02-02 & 8881.852 & 08:12:40.567 & 899.9977 & 20.95 & 0.4090±0.0008 & 0.1913±0.0014 & 0.5655±0.0116 & 0.3391±0.0010 & 0.4733±0.0024\\
        2020-02-03 & 8882.828 & 07:38:09.453 & 899.9978 & 21.95 & 0.4134±0.0008 & 0.1923±0.0014 & 0.6192±0.0112 & 0.3377±0.0010 & 0.4814±0.0024\\
        2020-02-04 & 8883.812 & 07:14:14.400 & 899.9978 & 20.70 & 0.4159±0.0008 & 0.1917±0.0014 & 0.5734±0.0115 & 0.3385±0.0010 & 0.4805±0.0026\\
        2020-02-05 & 8884.804 & 07:02:56.565 & 899.9977 & 17.75 & 0.4136±0.0010 & 0.1933±0.0017 & 0.5235±0.0120 & 0.3374±0.0012 & 0.4788±0.0030\\
        2020-02-07 & 8886.797 & 06:53:35.306 & 899.9977 & 21.15 & 0.4144±0.0008 & 0.1971±0.0015 & 0.6101±0.0110 & 0.3376±0.0010 & 0.4787±0.0026\\
        2020-02-09 & 8888.814 & 07:17:23.978 & 899.9977 & 22.70 & 0.4153±0.0008 & 0.1991±0.0014 & 0.6223±0.0103 & 0.3374±0.0010 & 0.4829±0.0023\\
        2020-02-14 & 8893.817 & 07:20:50.934 & 899.9977 & 21.45 & 0.4140±0.0008 & 0.1948±0.0014 & 0.5718±0.0110 & 0.3403±0.0010 & 0.4795±0.0025\\
        2020-02-18 & 8897.788 & 06:39:24.202 & 899.9977 & 20.60 & 0.4119±0.0008 & 0.1947±0.0015 & 0.5930±0.0115 & 0.3408±0.0010 & 0.4761±0.0026\\
        2020-02-20 & 8899.796 & 06:51:33.811 & 899.9977 & 21.60 & 0.4135±0.0008 & 0.1942±0.0015 & 0.6074±0.0124 & 0.3383±0.0010 & 0.4853±0.0026\\
        2020-02-23 & 8902.770 & 06:13:11.622 & 899.9976 & 21.40 & 0.4116±0.0008 & 0.1928±0.0015 & 0.5338±0.0116 & 0.3398±0.0010 & 0.4834±0.0027\\
        2020-03-02 & 8910.744 & 05:36:33.372 & 899.9978 & 19.85 & 0.4111±0.0009 & 0.1960±0.0016 & 0.6049±0.0121 & 0.3388±0.0011 & 0.4824±0.0028\\
        2020-03-04 & 8912.738 & 05:26:47.400 & 899.9978 & 17.10 & 0.4094±0.0010 & 0.1944±0.0018 & 0.6141±0.0144 & 0.3403±0.0013 & 0.4877±0.0033\\
        2020-03-07 & 8915.725 & 05:09:37.162 & 899.9978 & 20.50 & 0.4063±0.0009 & 0.1935±0.0018 & 0.5576±0.0220 & 0.3416±0.0012 & 0.4815±0.0031\\
        2020-03-08 & 8916.728 & 05:13:35.134 & 899.9978 & 26.10 & 0.4098±0.0007 & 0.1932±0.0014 & 0.5747±0.0103 & 0.3408±0.0009 & 0.4895±0.0024\\
        2020-03-09 & 8917.748 & 05:42:01.730 & 899.9978 & 21.65 & 0.4100±0.0009 & 0.1991±0.0017 & 0.5883±0.0120 & 0.3404±0.0011 & 0.4911±0.0029\\
        2020-03-16 & 8924.697 & 04:28:21.516 & 899.9980 & 25.40 & 0.4118±0.0007 & 0.1914±0.0014 & 0.6151±0.0110 & 0.3394±0.0009 & 0.4825±0.0024\\
        2020-03-18 & 8926.702 & 04:36:19.396 & 899.9980 & 26.20 & 0.4181±0.0007 & 0.1978±0.0013 & 0.7304±0.0118 & 0.3371±0.0009 & 0.4817±0.0024\\
        2020-03-20 & 8928.687 & 04:14:34.705 & 899.9981 & 24.55 & 0.4039±0.0007 & 0.1949±0.0014 & 0.6306±0.0118 & 0.3396±0.0009 & 0.4801±0.0025\\
        2020-03-20 & 8928.730 & 05:16:37.229 & 899.9981 & 22.80 & 0.4054±0.0008 & 0.1929±0.0015 & 0.6186±0.0128 & 0.3384±0.0010 & 0.4823±0.0027\\
        2020-03-22 & 8930.681 & 04:02:24.652 & 1199.9981 & 14.15 & 0.4070±0.0013 & 0.1992±0.0025 & 0.5142±0.0173 & 0.3412±0.0017 & 0.4755±0.0043\\
		\hline
	\end{tabular}%
	}
\end{table*}


\bsp	
\label{lastpage}

\clearpage

\includepdf[pages=-]{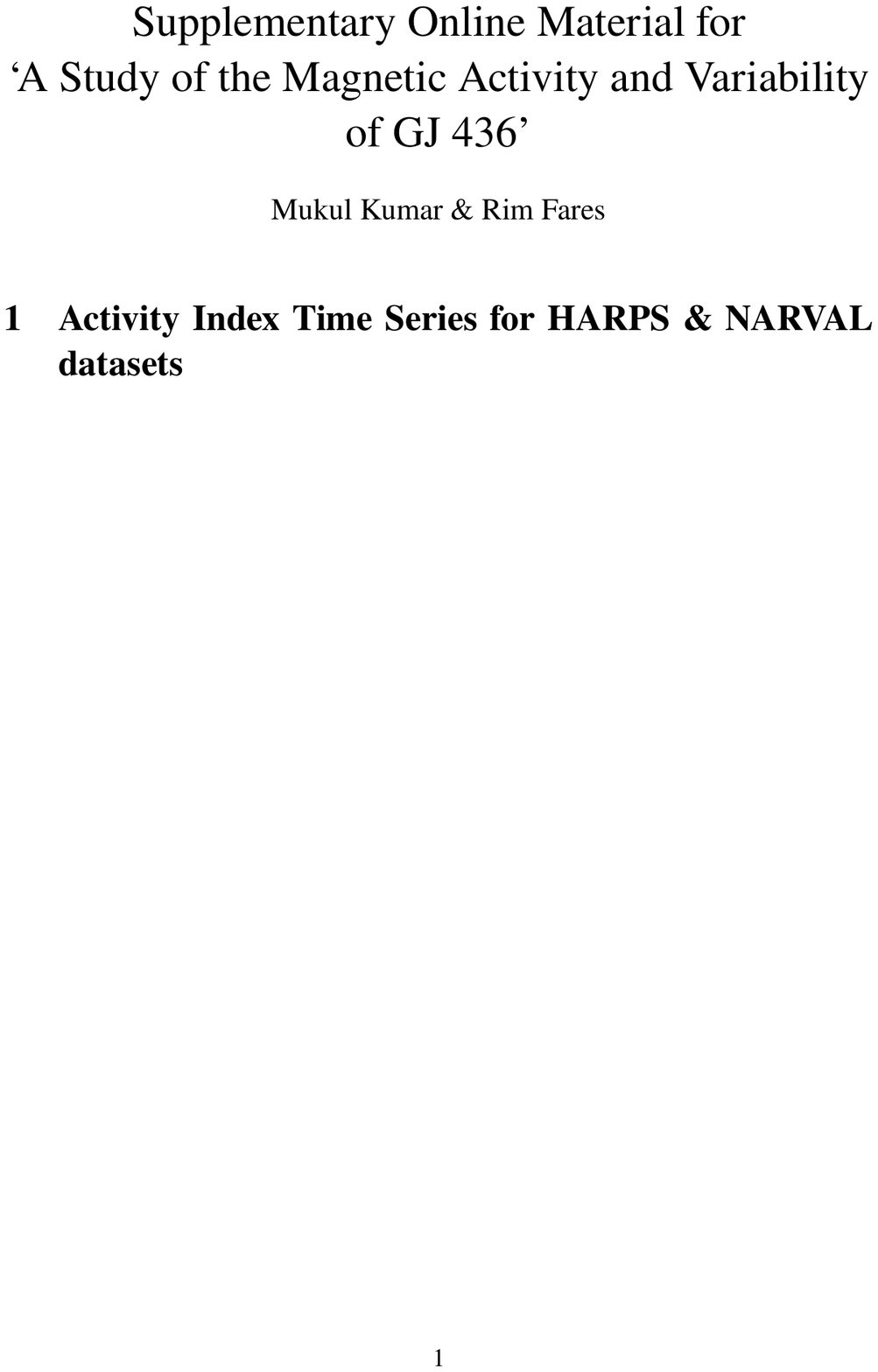}

\end{document}